\documentclass{revtex4}

\usepackage{amsmath}
\usepackage{amssymb}
\usepackage{amsfonts}
\usepackage{bm}
\usepackage{color}
\usepackage{graphicx}

\begin{document}

%\baselineskip 7mm

%------------------------------------------------------------------------------
% title
\title{\bf A Transient Bond Model for Dynamic Constraints in Meso-Scale
Coarse-Grained Systems}
\author{Takashi Uneyama}
\affiliation{Center for Computational Science,
Graduate School of Engineering, Nagoya University}

\date{\today}

\begin{abstract}
The dynamical properties of entangled polymers originate from the
dynamic constraints due to the uncrossability {between}
polymer chains.
We propose a highly coarse-grained simulation model with transient bonds
for such dynamically constrained systems. Based on the {ideas}
of the responsive particle {dynamics} (RaPiD) model [P. Kindt and W. J. Briels,
J. Chem. Phys. {\bf 127}, {134901} (2007)] and the multi-chain
slip-spring model [T. Uneyama and Y. Masubuchi, J. Chem. Phys. {\bf
137}, 154902 (2012)], we construct the RaPiD type transient bond model
as a coarse-grained slip-spring model. In our model,
{a polymer chain is expressed as a single particle, and
particles are connected by transient bonds.} The transient bonds
modulate the dynamics of particles but they do not affect {static}
properties {in equilibrium}.
We show the relation between {parameters for} the entangled polymer systems
and {those for} the transient bond model. By performing simulations based on the
transient bond model, we show how model parameters affect the linear
viscoelastic behavior and the diffusion behavior. We also show that the
viscoelastic behavior of entangled polymer systems can be well
reproduced by the transient bond model.
\end{abstract}

\maketitle

%------------------------------------------------------------------------------
% main text
%

\section{Introduction}

The entangled polymer systems exhibit characteristic relaxation
behavior such as very long relaxation time which strongly depends on
the degree of polymerization\cite{Doi-Edwards-book}.
{Because of their very long relaxation times, the simulations for polymer melts and solutions with large
degrees of polymerization are difficult. Especially, simulations for long time
relaxation relaxation processes by
microscopic molecular dynamics models (such as the Kremer-Grest
molecular dynamics model\cite{Kremer-Grest-1990}) are quite difficult.}
Instead of the microscopic models, mesoscopic
coarse-grained models have been proposed and utilized to study the long
time relaxation behavior of entangled {polymer} systems. Due to the
nature of the entanglement, however, the mesoscopic models are mainly
{constructed as} phenomenological dynamical models.
There are many mesoscopic phenomenological models which
reproduce characteristic relaxation behavior of entangled polymers.
The Doi-Edwards tube model\cite{Doi-Edwards-book} is based
on the tube picture in which the dynamics of an entangled polymer chain
is constrained by a tube like obstacle. The
slip-link\cite{Hua-Schieber-1998,Masubuchi-Takimoto-Koyama-Ianniruberto-Greco-Marrucci-2001,Schieber-2003,Doi-Takimoto-2003,Nair-Schieber-2006,Khaliullin-Schieber-2009}
and slip-spring\cite{Likhtman-2005,Uneyama-2011,Chappa-Morse-Zippelius-Muller-2012,Uneyama-Masubuchi-2012} 
models employ dynamic links which constrain the motion of polymer chains.
The properties of these mesoscopic models depend on the details of the
model, and various models and their properties have been studied.
Among various mesoscopic coarse-grained models, the multi-chain
slip-spring model has some interesting properties\cite{Uneyama-Masubuchi-2012}. The multi-chain
slip-spring model employs the slip-springs as the extra {thermodynamic} degrees of
freedom, and it has {a} well-defined (effective) free energy of the
system. Both the static and dynamic properties {of the
model} are designed to be
statistical mechanically sound, based on the (effective) free energy.

Most of the mesoscopic coarse-grained models employ the entanglement
{segment} as the basic kinetic unit.
{The coarse-graining levels of these coarse-grained models
are similar.}
Because the numerical costs in simulations
depend on the coarse-graining level,
highly coarse-grained models with much larger basic
kinetic {units} are preferred to perform simulations for
well-entangled systems with very long relaxation times.
Kindt and Briels proposed a highly coarse-grained model which is called
the responsive particle dynamics (RaPiD) model\cite{Kindt-Briels-2007}. In the RaPiD model, one
polymer chain is simply expressed by a single particle. Compared with
other mesoscopic models for entangled polymers, the characteristic
length scale of the RaPiD model is large and thus the RaPiD model is
numerically efficient. Due to its highly coarse-grained nature, unlike
other mesoscopic models, the entanglement effect cannot be directly
expressed in the RaPiD model by some mesoscopic objects such as tubes and
slip-links. Instead of the tubes and slip-links, Kindt and Briels introduced
the transient potential between particles. The transient potential
depends on the number of entanglements between two particles, and the
number of entanglements is treated as {the} extra degrees of freedom in the
system. In equilibrium, the equilibrium number of entanglements between
particles is {assumed to be} a function of the distance between two particles. Then, by
employing an effective free energy for the particle positions and the
numbers of entanglements, the dynamics of the system can be modeled by
the Langevin equations for the positions and the numbers of entanglements.
Kindt and Briels constructed the dynamic equations of the RaPiD model
and showed that the coarse-grained simulations with the RaPiD model can
successfully reproduce the dynamics of entangled polymers such as the
linear viscoelasticity. The RaPiD model has been extended to other
systems such as polymer solutions\cite{SantosdeOliveira-Fitzgerald-denOtter-Briels-2014}, associative telechelic polymers\cite{Sprakel-Spruijt-vanderGucht-Padding-Briels-2009,Sprakel-Padding-Briels-2011}, and
star polymers\cite{Liu-Padding-denOtter-Briels-2013,Liu-denOtter-Briels-2014,Fitzgerald-Lentzakis-Sakellariou-Vlassopoulos-Briels-2014,Fitzgerald-Briels-2018}, and has been shown to
be a useful coarse-grained model.

There are some similarities between the multi-chain slip-spring model
and the RaPiD model, although {their} coarse-graining levels are different.
Both models employ the extra degrees of freedom to express the
entanglement effect, and the effective free energies are utilized to
characterize the equilibrium probability distribution. From the similarity
between these models, we may {consider} the RaPiD model as a highly
coarse-grained version of the multi-chain slip-spring model, and unify
these models. In this
work, we propose a RaPiD type highly coarse-grained transient bond model{,}
based on the framework of the multi-chain slip-spring model. We show
that we can construct a highly coarse-grained model with the transient
bonds and the effective free energy. In our model, transient bonds
modulate the dynamics of particles but they do not affect the
equilibrium properties.
As an ideal case, we consider the system where
the equilibrium properties reduce to those of an ideal gas.
{This ideal version of the} transient bond model exhibits interesting
dynamical properties
{even while statically it is an ideal gas.}
We perform simulations for ideal transient bond systems and study how model
parameters affect the dynamic properties such as the diffusion and
linear viscoelasticity. 
We also consider the relation between {parameters for} the entangled
polymer systems and {those for} the ideal transient bond model.

\section{Model}

\subsection{Transient Bond Model}

Based on the idea by Kindt and Briels\cite{Kindt-Briels-2007}, we express one polymer chain by
one particle and introduce a transient potential between particles.
Unlike the original RaPiD model by Kindt and Briels, we
express the transient potential in terms of the discrete number of bonds
between particles.
We assume that the transient bonds modulate the dynamics of particles
but do not affect the equilibrium {probability distribution
of the particle positions}.
As we show in {what follows}, 
{the assumption that a number of transient bonds is discrete}
is convenient when we consider static, equilibrium properties.

We start from the equilibrium probability distribution without transient
bonds. We consider a three dimensional
system which consists of $M$ particles and has {a} volume $V$.
We express the position of the $i$-th particle is given as $\bm{R}_{i}$ ($i =
1,2,\dots,M$). We assume that the interaction between particles is
expressed by two-body interaction potential $v(\bm{r})$.
Then the partition function
can be simply expressed as
\begin{equation}
 \label{partition_function_equilibrium}
  \mathcal{Z} = \frac{1}{\Lambda^{3 M} M!} \int d\lbrace \bm{R}_{i}
  \rbrace \,
 \exp\left[ - \frac{1}{k_{B} T} \sum_{i > j} v(\bm{R}_{i} -
       \bm{R}_{j}) \right] ,
\end{equation}
where $\Lambda$ is the thermal de Broglie wavelength, $\int d\lbrace
\bm{R}_{i}\rbrace \dotsb$ represents the integral over all particle positions, $k_{B}$ is the Boltzmann constant, and
$T$ is the temperature.
The equilibrium probability distribution for the particle
positions is simply given as
\begin{equation}
 \label{probability_distribution_positions_equilibrium}
  P_{\text{eq}}(\lbrace \bm{R}_{i} \rbrace) = \frac{1}{\mathcal{Z}
  \Lambda^{3M} M!}
  \exp\left[ - \frac{1}{k_{B} T} \sum_{i > j} v(\bm{R}_{i} -
       \bm{R}_{j}) \right] .
\end{equation}

To express the dynamic constraint {effects}, we introduce the transient bonds to
the system. We express the number of bonds between the $i$-th and $j$-th
particles as $n_{ij}$. (In this model, we assume that multiple bonds can
share the same site without any penalties. In other words, we assume
transient bonds to be bosons. For convenience, we also
assume that bonds at the same site are {basically
indistinguishable}.)
Then, the state of the system can be expressed by
{two sets of} variables $\lbrace \bm{R}_{i} \rbrace$ and $\lbrace n_{ij}
\rbrace$. We express the transient interaction energy per bond as
$u(\bm{r})$ ($\bm{r}$ is the distance vector between particles). Then, the total transient interaction energy becomes
\begin{equation}
 \mathcal{U}_{\text{transient}}(\lbrace \bm{R}_{i} \rbrace,\lbrace n_{ij} \rbrace) = 
  \sum_{i > j} n_{ij} u(\bm{R}_{i} - \bm{R}_{j}) .
\end{equation}
We should introduce something to control the bond number. In this work,
we introduce the effective chemical potential. The idea of the effective chemical
potential was first introduced to a slip-link model for entangled polymers
by Schieber\cite{Schieber-2003}, and later applied to the multi-chain
slip-spring model\cite{Uneyama-Masubuchi-2012}.
We express the effective chemical potential for a transient bond as
$\mu$.

{The introduction of the transient bonds affect the
equilibrium probability distribution for particle positions, and
particles effectively feel an attractive potential.
This is the same as the case of the multi-chain slip-spring model, where
the slip-springs generate the effective attraction between
chains\cite{Uneyama-Horio-2011,Chappa-Morse-Zippelius-Muller-2012,Uneyama-Masubuchi-2012}.
Such an attractive interaction
is an artifact {of the model, and should be cancelled
so that transient bonds do not affect the equilibrium probability
distribution for particles as given by
eq~\eqref{probability_distribution_positions_equilibrium}.}
{For this purpose, we introduce} the repulsive compensation
potential to the system. We require the following condition for the joint probability
distribution of the particle position and the bond number:
\begin{equation}
 \label{condition_for_probability_distribution_particles_bonds_equilibrium}
 \sum_{\lbrace n_{ij} \rbrace} P_{\text{eq}}(\lbrace
  \bm{R}_{i}\rbrace, \lbrace n_{ij} \rbrace) = P_{\text{eq}}(\lbrace
  \bm{R}_{i}\rbrace),
\end{equation}
where $P_{\text{eq}}(\lbrace \bm{R}_{i}\rbrace)$ should be exactly the
same as one given by
eq~\eqref{probability_distribution_positions_equilibrium}, and
the summation over $n_{ij}$ is taken for all the possible bond numbers.
(Notice that,
in general, $P_{\text{eq}}(\lbrace \bm{R}_{i}\rbrace)$ can be different from
eq~\eqref{probability_distribution_positions_equilibrium} if we
introduce the bond potential\cite{Uneyama-Horio-2011}.)
}
Then, the
equilibrium conditional probability distribution {should} be expressed as follows:
\begin{equation}
 \label{probability_distribution_bonds_equilibrium}
 P_{\text{eq}}(\lbrace n_{ij} \rbrace | \lbrace \bm{R}_{i} \rbrace)
  = \frac{1}{\displaystyle \Xi(\lbrace \bm{R}_{i} \rbrace)
  \prod_{i > j} n_{ij}!}
  \exp\left[  \frac{\mu}{k_{B} T}\sum_{i > j} n_{ij}
      - \frac{1}{k_{B} T} \sum_{i > j} n_{ij} u(\bm{R}_{i} - \bm{R}_{j})
      \right] .
\end{equation}
where $\Xi(\lbrace \bm{R}_{i} \rbrace)$ is the grand partition function
under given particle positions.
(This grand partition is not the (full) grand partition
function of the system.
The grand partition function in
eq~\eqref{probability_distribution_bonds_equilibrium} should be interpreted
as the normalization factor.)
The explicit form of the grand partition function {becomes}
\begin{equation}
 \label{grand_partition_function_equilibrium}
  \begin{split}
 \begin{split}
    \Xi(\lbrace \bm{R}_{i} \rbrace) 
  & \, {= } \prod_{i > j} \sum_{n_{ij} = 0}^{\infty} \frac{1}{n_{ij}!}
  \exp\left[  \frac{\mu}{k_{B} T}n_{ij}
      - \frac{1}{k_{B} T} n_{ij} u(\bm{R}_{i} - \bm{R}_{j}) \right] \\
 & = \exp\left[ {\xi} \sum_{i > j} e^{- u(\bm{R}_{i} -
  \bm{R}_{j}) / k_{B} T} \right] ,
 \end{split}  
\end{split}
\end{equation}
with the effective fugacity (activity) $\xi \equiv e^{\mu / k_{B} T}$.

{The form of the compensation potential is automatically determined by
the condition~\eqref{condition_for_probability_distribution_particles_bonds_equilibrium}.}
From eqs~\eqref{probability_distribution_positions_equilibrium},
\eqref{probability_distribution_bonds_equilibrium}, and
\eqref{grand_partition_function_equilibrium},
the joint equilibrium probability distribution
{becomes} as follows:
\begin{equation}
 \begin{split}
 \label{probability_distribution_positions_bonds_equilibrium}
  P_{\text{eq}}(\lbrace \bm{R}_{i}
  \rbrace, \lbrace n_{ij} \rbrace) & =
 P_{\text{eq}}(\lbrace n_{ij} \rbrace |
 \lbrace \bm{R}_{i}  \rbrace) 
 P_{\text{eq}}(\lbrace \bm{R}_{i}
  \rbrace) \\
  & = \frac{1}{\displaystyle \mathcal{Z} \Lambda^{3 M} M!}
  \left[ {\prod_{i > j} \frac{\xi^{n_{ij}}}{n_{ij}!}} \right]
  \exp\Bigg[ 
      - \frac{1}{k_{B} T} \sum_{i > j} n_{ij} u(\bm{R}_{i} - \bm{R}_{j}) \\
  & \qquad       - \frac{1}{k_{B} T} \sum_{i > j} v(\bm{R}_{i} -
  \bm{R}_{j})
  - {\xi} \sum_{i > j} e^{- u(\bm{R}_{i} - \bm{R}_{j})
  / k_{B} T}\Bigg] .
 \end{split}
\end{equation}
The last term in the exponent of eq
\eqref{probability_distribution_positions_bonds_equilibrium}{,
which originates from the condition
~\eqref{condition_for_probability_distribution_particles_bonds_equilibrium}, } can be
interpreted as the {pairwise} repulsive compensation potential between particles.
{The compensation potential cancels the effective
attractive interaction by transient bonds, and recovers
eq~\eqref{probability_distribution_positions_equilibrium} when we take
the statistical average over the bond numbers.}
The effective
free energy of the system is defined {from
eq~\eqref{probability_distribution_positions_bonds_equilibrium}} as
\begin{equation}
 \label{effective_free_energy}
 \mathcal{F}(\lbrace \bm{R}_{i}
  \rbrace, \lbrace n_{ij} \rbrace)
  \equiv
   \sum_{i > j} v(\bm{R}_{i} - \bm{R}_{j}) 
  + \sum_{i > j} n_{ij} u(\bm{R}_{i} - \bm{R}_{j}) 
  + k_{B} T {\xi} \sum_{i > j} e^{- u(\bm{R}_{i} - \bm{R}_{j}) / k_{B} T} .
\end{equation}

To simulate dynamical properties, we need dynamic equations.
The dynamic equations should satisfy the detailed balance condition in
equilibrium. For the dynamics of the particle positions, we employ the
overdampled Langevin equation:
\begin{equation}
 \label{langevin_equation_particles}
 \frac{d\bm{R}_{i}(t)}{dt} = - \frac{1}{\zeta} \frac{\partial \mathcal{F}(\lbrace \bm{R}_{i}
  \rbrace, \lbrace n_{ij} \rbrace)}{\partial \bm{R}_{i}} +
  \bm{\kappa}(t) \cdot \bm{R}_{i} + \sqrt{\frac{2 k_{B} T}{\zeta}}
  \bm{w}_{i}(t) ,
\end{equation}
where $\zeta$ is the friction coefficient for a particle,
$\bm{\kappa}(t)$ is the velocity gradient tensor, and $\bm{w}(t)$ is the
Gaussian white noise. The first and second moments of the noise are
given as
\begin{equation}
 \langle \bm{w}_{i}(t) \rangle = 0, \qquad
 \langle \bm{w}_{i}(t) \bm{w}_{j}(t') \rangle = \delta_{ij} \delta(t - t') \bm{1},
\end{equation}
where $\langle \dots \rangle$ represents the statistical average and
$\bm{1}$ is the unit tensor. For the dynamics of the
{transient} bonds, we employ a
simple birth-death type dynamics\cite{vanKampen-book}. We assume that the transient bonds are
destroyed by a constant rate, and there is no direct correlation between
the destruction of different bonds. Then, we have the following transition rate
for the {decreasing} process (the destruction rate):
\begin{equation}
 \label{destruction_rate_bonds}
 W(n_{ij} - 1 | n_{ij}) = \frac{n_{ij}}{\tau},
\end{equation}
where $\tau$ is a constant which represents the average life time of
the transient bond. The transition rate for the {increasing} process (the construction rate) is automatically determined
from the detailed balance condition:
\begin{equation}
 \label{detailed_balance_condition_bonds}
 W({n_{ij} | n_{ij} + 1}) P_{\text{eq}}(n_{ij} + 1 | \lbrace \bm{R}_{i}
  \rbrace) = 
  {W(n_{ij} + 1| n_{ij})
  P_{\text{eq}}(n_{ij} | \lbrace \bm{R}_{i} \rbrace)} .
\end{equation}
From
eqs~{\eqref{probability_distribution_positions_bonds_equilibrium}}, \eqref{destruction_rate_bonds}, and
\eqref{detailed_balance_condition_bonds}, we have
\begin{equation}
 \label{construction_rate_bonds}
 W(n_{ij} + 1 | n_{ij}) =  \frac{1}{\tau} \xi e^{- u(\bm{R}_{i} -
 \bm{R}_{j}) / k_{B} T}.
\end{equation}
The dynamics of the system can be completely described by eqs
\eqref{langevin_equation_particles}, \eqref{destruction_rate_bonds}, and
\eqref{construction_rate_bonds}. We can perform dynamics simulations by
discretizing them and {solving} simulating time-evolution with some numerical schemes.

Before we proceed {to the analysis of the model}, we briefly comment on the relation
{between} our model {and} the original RaPiD model.
Our transient bond model may look {rather} different
from the standard RaPiD model. However, if the transient bond number is
large, we can approximate the bond number as a continuum variable. Then
we can show that our model reduces to the RaPiD type model, as shown in
Appendix~\ref{approximation_for_large_bond_number}.
Roughly speaking, the differences between our model and the original RaPiD model are the
factor {of the bond potential} in the effective free energy, and the friction
coefficients for particle positions and bond numbers.
Thus we believe that our
model can be utilized as a simplified version of the RaPiD model.

To calculate the viscoelastic properties, we need {an}
expression of the stress tensor. In many {cases},
the stress tensor {of a mesoscopic models for polymer} is assumed
to {obey} by the stress-optical rule. However, in the transient bond
model (and also in the RaPiD model), we do not have any
intra-chain, conformational degrees of freedom. {Thus a
naive application of the stress-optical rule does not work.}
Instead of the stress-optical
rule, we simply employ the Kramers form {stress tensor}:
\begin{equation}
 \label{stress_tensor}
 \begin{split}
  \hat{\bm{\sigma}}
  & = \frac{1}{V} \sum_{i > j} \frac{\partial \mathcal{F}(\lbrace \bm{R}_{i} \rbrace,\lbrace n_{ij}
  \rbrace)}{\partial (\bm{R}_{i} - \bm{R}_{j})} (\bm{R}_{i} -
  \bm{R}_{j}) - \frac{M k_{B} T}{V} \bm{1} \\
  & = \frac{1}{V} \sum_{i > j}
  \left[ \left[ n_{ij}   - \xi e^{- u(\bm{R}_{i} - \bm{R}_{j}) / k_{B} T} \right] \frac{\partial u(\bm{R}_{i} - \bm{R}_{j}) }{\partial
  (\bm{R}_{i} - \bm{R}_{j})}
  + \frac{\partial v(\bm{R}_{i} - \bm{R}_{j}) }{\partial (\bm{R}_{i} - \bm{R}_{j})}
  \right] (\bm{R}_{i} - \bm{R}_{j})
  - \frac{M k_{B} T}{V} \bm{1} .
 \end{split}
\end{equation}
In the slip-spring model, the stress tensor consists of two parts; the
real stress and the virtual stress\cite{Ramirez-Sukumaran-Likhtman-2007,Uneyama-2011,Uneyama-Masubuchi-2012}. The real stress comes from the bond
vectors and is consistent with the stress-optical rule\cite{Doi-Edwards-book,Inoue-Osaki-1996}. The virtual
stress comes from the slip-springs, and sometimes it is neglected
because it is (at least apparently) not consistent with the
stress-optical rule.
Whether we should include the contribution of the
virtual stress or not is not trivial\cite{Uneyama-2011,Uneyama-Masubuchi-2012}. In some cases, the virtual stress
term is simply discarded {to recover the stress-optical
rule. (In the slip-spring model, the contribution of the virtual stress
seems not to be so important, at least qualitatively\cite{Uneyama-2011,Uneyama-Masubuchi-2012}.)}
In our model, due to the lack of the intra-chain
degrees of freedom, we need to include the virtual stress, otherwise the
effect of the transient bonds to the viscoelastic properties will be
ignored.
{
Although eq~\eqref{stress_tensor} is not (at least apparently)
consistent with the stress-optical rule,
we should mention that the stress-optical rule is an empirical rule and
it does not hold under some situations such as under fast extensional
flows\cite{Kroger-Luap-Muller-1997}. The simulation data by the
Kremer-Grest model showed that the contributions from the non-bonded
interaction to the total stress is not simple\cite{Ramirez-Sukumaran-Likhtman-2007}.}

The time evolution of the system can be formally expressed by the Fokker-Planck
operator\cite{vanKampen-book} associated with the Langevin equation
(eq \eqref{langevin_equation_particles}) and the reconstruction rates
(eqs \eqref{destruction_rate_bonds} and
\eqref{construction_rate_bonds}). Eq \eqref{langevin_equation_particles}
contains the contribution of the external flow, which is absence in
equilibrium, and thus we decompose the time evolution operator into the
equilibrium and flow {parts}:
\begin{equation}
 \label{time_evolution_equation}
 \frac{\partial P(\lbrace \bm{R}_{i} \rbrace,\lbrace n_{ij}
  \rbrace,t)}{\partial t} 
  = [\mathcal{L}_{0} + \mathcal{L}_{1}(t) ] P(\lbrace \bm{R}_{i} \rbrace,\lbrace n_{ij}
  \rbrace,t) .
\end{equation}
Here, $\mathcal{L}_{0}$ is the equilibrium time-evolution operator
(which is the sum of the Fokker-Planck operator in equilibrium and the
transition matrices), and $\mathcal{L}_{1}(t)$ is the time-evolution
operator by the external flow:
\begin{equation}
 \label{time_evolution_operator_flow}
 \mathcal{L}_{1}(t) P = - \sum_{i} \frac{\partial}{\partial \bm{R}_{i}} \cdot
  [\bm{\kappa}(t) \cdot \bm{R}_{i} P ] .
\end{equation}
If {the contribution of} the external flow is sufficiently small, the time evolution operator
by {the} flow{, $\mathcal{L}_{1}(t)$,} can be interpreted as
{a} perturbation. Then we can expand the
probability distribution into the {perturbation} series to
consider the linear
response:
\begin{equation}
 \label{probability_distribution_perturbation_expansion}
 P(\lbrace \bm{R}_{i} \rbrace,\lbrace n_{ij}
  \rbrace,t) = P_{\text{eq}}(\lbrace \bm{R}_{i} \rbrace,\lbrace n_{ij}
  \rbrace) + P_{1}(\lbrace \bm{R}_{i} \rbrace,\lbrace n_{ij}
  \rbrace,t) .
\end{equation}
By substituting eqs \eqref{time_evolution_operator_flow}
\eqref{probability_distribution_perturbation_expansion} into eq
\eqref{time_evolution_equation}, and retaining only the leading order
terms, we have
\begin{equation}
 \label{time_evolution_equation_perturbation_expansion}
\begin{split}
 \frac{\partial P_{1}(\lbrace \bm{R}_{i} \rbrace,\lbrace n_{ij}
  \rbrace,t)}{\partial t}
 & \approx \mathcal{L}_{1}(t)  P_{\text{eq}}(\lbrace \bm{R}_{i}
  \rbrace,\lbrace n_{ij}  \rbrace)+ \mathcal{L}_{0} P_{1}(\lbrace \bm{R}_{i} \rbrace,\lbrace n_{ij}
  \rbrace,t) \\
 & = \frac{V}{k_{B} T} \hat{\bm{\sigma}} : \bm{\kappa}(t) P_{\text{eq}}(\lbrace \bm{R}_{i}
  \rbrace,\lbrace n_{ij}  \rbrace)+ \mathcal{L}_{0} P_{1}(\lbrace \bm{R}_{i} \rbrace,\lbrace n_{ij}
  \rbrace,t) .
\end{split}
\end{equation}
Here we have utilized $\mathcal{L}_{0} P_{\text{eq}} = 0$. Eq
\eqref{time_evolution_equation_perturbation_expansion} can be formally
integrated as
%\begin{equation}
% P_{1}(\lbrace \bm{R}_{i} \rbrace,\lbrace n_{ij}
%  \rbrace,t)
%  = \int_{-\infty}^{t} dt' \, e^{(t - t') \mathcal{L}_{0}}
%- \sum_{i} \frac{\partial}{\partial \bm{R}_{i}} \cdot
%  \bm{\kappa}(t') \cdot \bm{R}_{i}  P_{\text{eq}}(\lbrace \bm{R}_{i} \rbrace,\lbrace n_{ij}
%  \rbrace)
%\end{equation}
\begin{equation}
 \label{time_evolution_equation_perturbation_solution}
 P_{1}(\lbrace \bm{R}_{i} \rbrace,\lbrace n_{ij}
  \rbrace,t)
  = \frac{V}{k_{B} T} \int_{-\infty}^{t} dt' \, e^{(t - t') \mathcal{L}_{0}}
 \hat{\bm{\sigma}} :
  \bm{\kappa}(t')   P_{\text{eq}}(\lbrace \bm{R}_{i} \rbrace,\lbrace n_{ij}
  \rbrace) ,
\end{equation}
and the ensemble average of the stress tensor at time $t$ is calculated from
eq \eqref{time_evolution_equation_perturbation_solution}:
\begin{equation}
 \label{stress_tensor_perturbation_solution}
 \begin{split}
  \bm{\sigma}(t) 
  & = \int d\lbrace \bm{R}_{i} \rbrace \sum_{\lbrace n_{ij} \rbrace} 
  \hat{\bm{\sigma}} [P_{\text{eq}}(\lbrace \bm{R}_{i} \rbrace,\lbrace n_{ij}
  \rbrace) + P_{1}(\lbrace \bm{R}_{i} \rbrace,\lbrace n_{ij}
  \rbrace,t) ]\\
  & = \langle \hat{\bm{\sigma}} \rangle_{\text{eq}} + \frac{V}{k_{B} T}
  {\int_{-\infty}^{t} dt'} \, \left\langle
  \hat{\bm{\sigma}}(t - t') \hat{\bm{\sigma}} \right\rangle_{\text{eq}}: \bm{\kappa}({t'}) ,
 \end{split}
\end{equation}
where $\langle \dots \rangle_{\text{eq}}$ represents the equilibrium
ensemble average, and $\hat{\bm{\sigma}}(t - t') \equiv e^{(t - t') \mathcal{L}_{0}^{\dagger}}
\hat{\bm{\sigma}}$ {($\mathcal{L}_{0}^{\dagger}$ is the adjoint operator
for $\mathcal{L}_{0}$)} is the time-shifted stress tensor. Eq
\eqref{stress_tensor_perturbation_solution} means that the relaxation
modulus tensor is given as the equilibrium auto-correlation function of the stress
tensor. For example, from eq
\eqref{stress_tensor_perturbation_solution}, the shear relaxation
modulus $G(t)$ becomes
\begin{equation}
 \label{shear_relaxation_modulus_formula}
 G(t) = \frac{V}{k_{B} T} \langle \hat{\sigma}_{xy}(t) \hat{\sigma}_{xy}
  \rangle_{\text{eq}} .
\end{equation}
This is consistent with the well-known Green-Kubo formula in the
linear response theory\cite{Evans-Morris-book}. (In the slip-spring model, the simple Green-Kubo
type form does not hold {when} we ignore the contribution of the virtual
stress\cite{Ramirez-Sukumaran-Likhtman-2007,Uneyama-2011,Uneyama-Masubuchi-2012}.)

\subsection{Ideal Transient Bond Model}

So far, the interaction potential $v(\bm{r})$ and the transient bond
potential $u(\bm{r})$ are rather arbitrary. In
this work, to investigate dynamical properties of the transient bond
model, we set the interaction potential $v(\bm{r})$ to zero and employ a simple harmonic bond potential for $u(\bm{r})$:
\begin{align}
 \label{zero_interaction_potential}
 v(\bm{r}) & = 0, \\
 \label{harmonic_bond_potential}
 u(\bm{r}) & = \frac{3 k_{B} T}{2 a^{2}} \bm{r}^{2} ,
\end{align}
where $a$ is the average bond size. In addition, we limit ourselves to
equilibrium systems and set $\bm{\kappa}(t) = 0$. 

We may call the transient
bond model with eqs \eqref{zero_interaction_potential} and
\eqref{harmonic_bond_potential} as the ideal transient bond model.
Eq \eqref{zero_interaction_potential} means that,
statically, our system is just an ideal gas {from the view point of the
static properties in equilibrium}. 
In fact, the partition function of the system {is} simply calculated to be 
\begin{equation}
 \mathcal{Z} = \frac{V^{M}}{\Lambda^{3 M} M!} ,
\end{equation}
and this gives the equation of state of an ideal gas.
(The contributions from the transient bonds and compensation potentials
cancel each other.)
Therefore, all the static properties of this ideal transient model
coincide to those of the ideal gas. For example, the pressure of the
ideal transient model in equilibrium is
$P = M k_{B} T / V$.
Of course, this does not mean that
the dynamical properties of the ideal transient model coincide to the
ideal gas. The dynamics of the particles is largely affected by the
transient bonds, and thus the dynamical properties such as the
viscoelasticity of the ideal transient bond model become qualitatively
different from those of the ideal gas without any transient interactions. This
would be clear from the Langevin equation
\eqref{langevin_equation_particles}, where the forces from the bond
and compensation potentials do not cancel.

By substituting eq
\eqref{harmonic_bond_potential} into eq \eqref{effective_free_energy},
we have the following simple effective free energy:
\begin{equation}
 \label{effective_free_energy_ideal}
  \frac{\mathcal{F}(\lbrace \bm{R}_{i}
  \rbrace, \lbrace n_{ij} \rbrace)}{k_{B} T}
  = \frac{3}{2 a^{2}} \sum_{i > j} n_{ij} (\bm{R}_{i} - \bm{R}_{j})^{2}
  + \xi \sum_{i > j} e^{- 3 (\bm{R}_{i} - \bm{R}_{j})^{2} / 2 a^{2}} .
\end{equation}
Also, the stress tensor (eq \eqref{stress_tensor}) is simplified as
\begin{equation}
 \label{stress_tensor_ideal}
  \hat{\bm{\sigma}} = 
  \frac{3 k_{B} T}{2 a^{2} V} \sum_{i > j}
  \left[ n_{ij}   - \xi e^{- 3 (\bm{R}_{i} - \bm{R}_{j})^{2} / 2
  a^{2}} \right] (\bm{R}_{i} - \bm{R}_{j}) (\bm{R}_{i} - \bm{R}_{j})
  - \frac{M k_{B} T}{V} \bm{1} .
\end{equation}
From eqs \eqref{probability_distribution_positions_bonds_equilibrium} and \eqref{stress_tensor_ideal}, the average stress tensor in equilibrium becomes
\begin{equation}
 \label{stress_tensor_ideal_equilibrium}
   \langle \hat{\bm{\sigma}} \rangle_{\text{eq}}
    = \int d\lbrace \bm{R}_{i} \rbrace \sum_{\lbrace n_{ij} \rbrace}
   \hat{\bm{\sigma}} P_{\text{eq}}(\lbrace n_{ij} \rbrace | \lbrace \bm{R}_{i}
   \rbrace)
   P_{\text{eq}}(\lbrace \bm{R}_{i} \rbrace) 
    =  - \frac{M k_{B} T}{V} \bm{1} .
\end{equation}
As expected, the contribution of the transient bonds
to the stress tensor {vanishes} in equilibrium{, and the
stress tensor simply consists of the pressure of the ideal gas.}

{We can analytically calculate} some properties of the ideal
transient model. The average number of transient bonds per pair is calculated as
\begin{equation}
 \begin{split}
  \langle n_{ij} \rangle_{\text{eq}}
  & = \int d\bm{R}_{i} d\bm{R}_{j} \sum_{n_{ij}} n_{ij}
  P_{\text{eq}}(\bm{R}_{i},\bm{R}_{j},n_{ij}) \\
  & = \int d\bm{R}_{i} d\bm{R}_{j} \frac{1}{V^{2}} \sum_{n_{ij} = 0}^{\infty}
  \frac{n_{ij}}{n_{ij}!}
 \xi^{n_{ij}}
 \exp\left[ - \frac{3 n_{ij} (\bm{R}_{i} - \bm{R}_{j})^{2}}{2 a^{2}}
      - \xi e^{-3 (\bm{R}_{i} - \bm{R}_{j})^{2} / 2 a^{2}}
     \right] \\
%  & = \int d\bm{Q} \, \frac{1}{V} 
%  \xi e^{- 3 \bm{Q}^{2} / 2 a^{2}} \\
  & = \frac{\xi }{V}
  \left(\frac{2 \pi a^{2}}{3}\right)^{3/2} .
 \end{split}
\end{equation}
Thus the transient bond density (the number of transient bonds per unit
volume), $\phi$, is given as
\begin{equation}
 \label{average_bond_density}
 \phi = \frac{M (M - 1)}{2 V} \langle n_{ij} \rangle_{\text{eq}}
%  = \rho^{2} \xi \frac{(1 - 1 / M)}{2} 
%  \left(\frac{2 \pi a^{2}}{3}\right)^{3/2}
  \approx \frac{\xi \rho^{2}}{2}
  \left(\frac{2 \pi a^{2}}{3}\right)^{3/2} ,
\end{equation}
where $\rho = M / V$ is the number density of particles, and we have
assumed that the number of particles in the system is sufficiently large
($M \gg 1$). Eq \eqref{average_bond_density} is consistent with the
expression for the inter-chain slip-spring density in the multi-chain
slip-spring model\cite{Uneyama-Masubuchi-2012}.
Eq \eqref{average_bond_density} means that the transient
bond density is proportional to the effective fugacity $\xi$ and the
square of the density $\rho$. We can utilize both $\xi$ and $\rho$ to
control the transient bond density. (This situation is similar to but much
simpler than the multi-chain slip-spring model.)
The elastic modulus of the system at
the short time scale, $G_{0}$, is roughly proportional to the transient bond
density, $G_{0} {\propto \phi} \propto \xi \rho^{2}$.

In the ideal transient bond model, we have only three dimensionless parameters which
characterize {a  target} system; the number density of particles $\rho a^{3}$, the
effective fugacity $\xi$, and the {(relative)} average life time of the transient
bond {$\tau / \tau_{\text{TB}}$ (with $\tau_{\text{TB}}
\equiv a^{2} \zeta / k_{B} T$ being the characteristic time scale of the
diffusive motion of a particle)}.
For the sake of simplicity, we
employ dimensionless units by setting $a = 1$, $k_{B} T = 1$, and
$\zeta$ = 1. Then, the dynamical properties of the system can be fully
specified by the set of three parameters, $\rho$, $\xi$, and $\tau$. In
what follows we use the dimensionless units and study the effects of
these three parameters.

\subsection{Parameters for Entangled Polymer Systems}
\label{parameters_for_entangled_polymer_systems}

We consider the relation between the degree of polymerization (or the
molecular weight) and the dimensionless parameters in the ideal transient bond model, $\rho$, $\xi$, and $\tau$.
Because one polymer chain is expressed as a single particle, the
characteristic length scale of the ideal transient bond model depends on
the degree of polymerization.
Due to this property, the relation between {an} entangled polymer system and
the ideal transient bond model is not trivial.

We consider entangled polymer systems where the number of segment
density, $\rho_{\text{segment}}$, is constant and only the number of
segments (the degree of
polymerization), $N$, is changed. We express the size of {a} segment as
$b$. The particle density in the transient bond model can be interpreted as
the polymer chain density, and is simply expressed as
\begin{equation}
 \rho = \frac{\rho_{\text{segment}}}{N} .
\end{equation}
In the ideal transient bond model, we need the dimensionless particle density $\rho a^{3}$
rather than the {(dimensioned)} particle density itself. Thus we
need {to specify} the average length {of the bond} $a$. Unfortunately the
explicit form of $a$ in terms of polymer parameters is not clear. The
transient bonds originate from the entanglement effect, and the
entanglement effect becomes relevant only if two chains are overlapped.
Then, $a$ should depend on the radius of gyration of a polymer or the average
end-to-end {distance} of a polymer. Fortunately, both of these length scales
are proportional to $N b^{2}$. {Therefore it would be
reasonable to consider}
\begin{equation}
 a^{2} \propto N b^{2} .
\end{equation}
{In many cases, the unit length scale of a model is taken
to be $b$. However, in our model, {we should employ $a$ as }the unit length scale,
because a polymer chain is expressed as a single particle.}
Therefore, the unit length scale depends on the {degree of polymerization}
rather strongly, $a \propto N^{1/2}$. In other words, the degree of
coarse-graining depends on the degree of polymerization rather
strongly. {This $N$-dependent degree of coarse-graining
makes the relation between parameters in the transient bond model and
{an} entangled polymer system rather complicated. Here we consider
the $N$-dependence of several physical quantities to establish the
relation between {transient bond} model parameters and the degree of polymerization
$N$.}

The dimensionless particle density is given as
\begin{equation}
 \rho a^{3} 
  \propto \frac{\rho_{\text{segment}}}{N} N^{3/2} b^{3}
  \propto N^{1/2} .
\end{equation}
{Thus the particle density in the transient bond model
should be changed if we change the degree of polymerization.}
The transient bond density can be interpreted as the entanglement
density. The {(dimensioned)} entanglement density is constant if the polymers are
sufficiently long and the segment density is constant, $\phi \propto N^{0}$.
On the other hand, from eq \eqref{average_bond_density}, the dependence of the bond density to the segment
number is given as 
\begin{equation}
 \label{scale_transform_average_bond_density}
 \phi = \frac{\xi \rho^{2}}{2}
  \left(\frac{2 \pi a^{2}}{3} \right)^{3/2}
  \propto \frac{\xi \rho_{\text{segment}}^{2}}{N^{2}} N^{3/2} b^{3}
  \propto \xi N^{-1/2} .
\end{equation}
{To satisfy the condition $\phi \propto N^{0}$ with
eq~\eqref{scale_transform_average_bond_density}, the effective fugacity
should depend on $N$.}
Thus we have the following relation for the effective fugacity.
\begin{equation}
 \xi \propto N^{1/2} .
\end{equation}
The characteristic bond life time $\tau$ is roughly estimated to
be the same as the (pure)
reptation time $\tau_{\text{rep}} \propto N^{3}$. Also, the
{(dimensioned)} friction coefficient for a particle is 
related to the friction coefficient for the Rouse chain which moves
along the tube, ${\zeta \propto} N \zeta_{\text{segment}}$ ($\zeta_{\text{segment}}$ is
the friction coefficient of a segment). Thus we have {the
following relation for the dimensionless bond life time $\tau / \tau_{\text{TB}}$:}
\begin{equation}
 \label{scale_transform_life_time}
 {\frac{\tau}{\tau_{\text{TB}}}} =
 \frac{k_{B} T \tau}{a^{2} \zeta} \propto
  \frac{k_{B} T \tau_{\text{rep}}}{N b^{2} \times N \zeta_{\text{segment}} } \propto
  N .
\end{equation}

From the results {shown} above, all the three parameters in
our {transient bond} model can be
approximately related to the degree of polymerization $N$. If we have a
reference parameter set in the dimensionless units ($a = 1$, $k_{B} T =
1$, and $\zeta = 1$) as $\rho_{0}, \xi_{0}$, and $\tau_{0}$, for the
{reference} degree of polymerization $N_{0}$, we have
\begin{equation}
 \label{scale_transforms_for_entangled_polymers}
 \rho = \left(\frac{N}{N_{0}}\right)^{1/2} \rho_{0}, \qquad
 \xi = \left(\frac{N}{N_{0}}\right)^{1/2} \xi_{0}, \qquad
 \tau = \left(\frac{N}{N_{0}}\right)^{1} \tau_{0},
\end{equation}
for {a} system with the degree of polymerization $N$.
To convert the dimensionless units in the transient bond model to the
standard units {for an} entangled polymer system, we need the expressions of
the dimensionless units in the ideal transient bond model. The unit energy
$k_{B} T$ is common for {two models}, and thus we need only
the length and time units, {$a$ and $\tau_{\text{TB}}$}:
\begin{equation}
 \label{conversion_from_transient_bond_to_entangled_polymers}
  a \propto N^{1/2} b \propto N^{1/2}, \qquad
  \tau_{\text{TB}} = \frac{a^{2} \zeta}{k_{B} T} \propto \frac{N^{2} b^{2}
  \zeta_{\text{segment}}}{k_{B} T} \propto N^{2}.
\end{equation}
{
From~\eqref{conversion_from_transient_bond_to_entangled_polymers}, we
find that
the $N$-dependence of the characteristic time unit $\tau_{\text{TB}}$ is the same as
one of the Rouse time, $\tau_{R} \propto N^{2}$. Thus we may
interpret that the coarse-graining in our model is performed so that
the characteristic time scale becomes the Rouse time. This is consistent
with eq~\eqref{scale_transform_life_time} where we have
apparently weak $N$-dependence for $\tau$.}
Eqs \eqref{scale_transforms_for_entangled_polymers} and
\eqref{conversion_from_transient_bond_to_entangled_polymers} give only
the power law exponents for the degree of polymerization, and they do
not tell us about the numerical prefactors. To map the
results of the transient bond model to other mesoscopic models for
entangled polymers, we should phenomenologically determine the
numerical scale conversion factors. Once the scale conversion factors
are determined for one system, it is straightforward to obtain the
scale conversion factors for other systems with different degrees of polymerization.

\subsection{Numerical Scheme}

To perform simulations, 
we discretize time $t$ by the step size $\Delta t$, and set $t_{k} = k
\Delta t$.
We employ the stochastic Runge-Kutta method\cite{Honeycutt-1992} to integrate the Langevin
equation (eq \eqref{langevin_equation_particles}).
In the stochastic Runge-Kutta scheme, the update of the
position from time $t_{k}$ to $t_{k + 1}$ obeys:
\begin{align}
 \label{langevin_equation_particles_discretized_srk_1st}
 \bm{R}_{i,k}^{*} & = \bm{R}_{i,k} + \Delta t \, \bm{F}_{i}(\lbrace
 \bm{R}_{i,k} \rbrace) + \sqrt{2
 \Delta t} \, \bm{\theta}_{i,k} , \\
 \label{langevin_equation_particles_discretized_srk_2nd}
 \bm{R}_{i,k + 1} & = \bm{R}_{i,k} + \frac{\Delta t}{2}
 [\bm{F}_{i}(\lbrace \bm{R}_{i,k} \rbrace)
 + \bm{F}_{i}(\lbrace \bm{R}_{i,k}^{*} \rbrace)]
 + \sqrt{2 \Delta t} \, \bm{\theta}_{i,k}  ,
\end{align}
where $\bm{R}_{i,k} \equiv \bm{R}_{i}(t_{k})$, $\bm{F}_{i}$ is the force
acting on the $i$-th particle, and $\bm{\theta}_{i,k}$ is the Gaussian
white noise.
{The force is calculated as}
\begin{equation}
 \label{force_discretized}
 \bm{F}_{i}(\lbrace \bm{R}_{i} \rbrace)
  \equiv - 3 \sum_{j}
  \left[ n_{ij}
   - \xi e^{- 3(\bm{R}_{i} - \bm{R}_{j})^{2} / 2}  \right] (\bm{R}_{i} -
  \bm{R}_{j}) ,
\end{equation}
{and} the noise {is a Gaussian noise which is
generated to satisfy} the
following relations:
\begin{equation}
 \langle \bm{\theta}_{i,k} \rangle = 0, \qquad
 \langle \bm{\theta}_{i,k} \bm{\theta}_{j,l} \rangle = \delta_{ij}
 \delta_{kl} \bm{1} .
\end{equation}

We handle each transient bond separately, rather than directly handle
the number of the transient bond $n_{ij}$. From eq
\eqref{destruction_rate_bonds}, a transient bond connected to
the $i$-th and $j$-th particles will be destroyed by the destruction
rate $1 / \tau$. The construction rate (eq
\eqref{construction_rate_bonds}) is not changed even if we handle
existing transient bonds separately.
Then, we integrate the construction and destruction rates
(eq~\eqref{destruction_rate_bonds} {and $1 / \tau$}) from
$t_{k}$ to $t_{k + 1}$, to obtain the destruction probability for an existing bond and the
construction probability for a new bond connected to the $i$-th and $j$-th particles:
\begin{align}
 \label{destruction_probability_bonds}
 \Psi_{-} & = 1 - \exp(- \Delta t / \tau) ,\\
 \label{construction_probability_bonds}
 \Psi_{+,ij} &  = 1 - \exp\left[ - \xi e^{- 3 (\bm{R}_{i} -
 \bm{R}_{j})^{2} / 2} \Delta t / \tau \right] .
\end{align}
Since eq \eqref{construction_probability_bonds} decays rapidly as the
distance between particles increases, we assume that the
construction of bonds occurs only for pairs of which distance is rather
short. Therefore, practically, the construction trials are required only for the
pairs whitin a cut-off distance.

The numerical scheme for the dynamics simulation based on the transient
bond model is summarized as follows:
\begin{enumerate}
 \item \label{initialization}
       Initialize the particle positions and bonds. The particle
       positions are sampled from the uniform distribution. The
       bond numbers are sampled from the Poisson distribution{, for each pair}.
 \item \label{cell_list_construction}
       Construct a cell-list for the calculation of the force by
       eq~{\eqref{force_discretized}} and the destruction probability by eq
       \eqref{destruction_probability_bonds}.
 \item \label{integration_langevin}
       Integrate the Langevin equation for the positions, by the stochastic Runge-Kutta scheme
       (eqs \eqref{langevin_equation_particles_discretized_srk_1st} and
       \eqref{langevin_equation_particles_discretized_srk_2nd}). The
       force is calculated by eq~{\eqref{force_discretized}} only for the pairs within a cut-off distance.
 \item \label{sampling_destruction}
       Destroy the existing bonds by the destruction probability (eq
       \eqref{destruction_probability_bonds}).
 \item \label{sampling_construction}
       Construct new bonds by the construction probability (eq
       \eqref{construction_probability_bonds}). The construction trials
       are performed only for the pairs within a cut-off distance.
 \item Go to \ref{cell_list_construction}. and iterate the time evolution.
\end{enumerate}
It should be noted here that our numerical scheme shown above is similar to but much simpler than one for the
multi-chain slip-spring model\cite{Uneyama-Masubuchi-2012}. In the
slip-spring model, we need to stochastically hop {(slide)}
the slip-springs {on polymer chains}. We
also need to stochastically sample segments around the chain ends for
efficient calculations. {These are absent in the transient
bond model.}

During the simulation, the snapshots of the particle positions and the
stress tensor of the system are saved. We calculate the average
mean-square displacements of particles from the snapshots of particle
positions, to study the diffusion behavior. Also, we calculate the shear
relaxation moduli by
eq~\eqref{shear_relaxation_modulus_formula}{, to study the
viscoelastic behavior}.
To improve the statistical accuracy, we utilize the Likhtman's formula\cite{Likhtman-chapter}
instead of eq \eqref{shear_relaxation_modulus_formula}:
\begin{equation}
 \label{shear_relaxation_modulus_formula_likhtman}
\begin{split}
  G(t) & = \frac{V}{5 k_{B} T}
  [\langle \hat{\sigma}_{xy}(t) \hat{\sigma}_{xy} \rangle_{\text{eq}}
  + \langle \hat{\sigma}_{yz}(t) \hat{\sigma}_{yz} \rangle_{\text{eq}}
  + \langle \hat{\sigma}_{xz}(t) \hat{\sigma}_{zx} \rangle_{\text{eq}}
  ] \\
 & \qquad + \frac{V}{30 k_{B} T}
  [ \langle \hat{N}_{xy}(t) \hat{N}_{xy} \rangle_{\text{eq}}
  + \langle \hat{N}_{yz}(t) \hat{N}_{yz} \rangle_{\text{eq}}
  + \langle \hat{N}_{zx}(t) \hat{N}_{zx} \rangle_{\text{eq}}
  ] ,
\end{split}
\end{equation}
where $\hat{N}_{\alpha \beta} = \hat{\sigma}_{\alpha\alpha} - \hat{\sigma}_{\beta\beta} $
is the normal stress difference.

\section{Results}
\label{results}

\subsection{Effect of Dimensionless Parameters}
\label{effect_of_dimensionless_parameters}

We perform simulations for different {dimensionless} parameter sets. To study the
effects of individual parameters to the dynamical properties, we employ a
parameter set $\rho = 1$, $\xi = 1$, and $\tau = 100$ as a
reference parameter set, and change one parameter systematically. We
change $\rho$ as $\rho = 0.125, 0.25, 0.5, 1$, and $2$ for $\xi = 1$ and $\tau
= 100$, and change $\xi$ as $\xi = 0.125, 0.25, 0.5, 1$, and $2$ for $\rho =
1$ and $\tau = 100$, and change $\tau$ as $\tau = 0.1, 1, 10, 100$, and $1000$
for $\rho = 1$ and $\xi = 1$. The system size is taken to be
sufficiently large (typically $V = 8^{3}$) and the periodic boundary
condition is applied to all the directions.
The time step size is set
to be $\Delta t = 0.01$. The cut-off distance $r_{\text{cut}}$ is set so
that $\xi e^{- 3 r_{\text{cut}}^{2} / 2} = 10^{-4}$. (Thus the cut-off
distance depends on the value of $\xi$.)
We perform equilibrium simulations {with different random
seeds} for the same parameter set, and then take
averages over time and different samples to improve the statistical
accuracy. (We utilize the Mersenne twister
method\cite{Matsumoto-Nishimura-1998} to generate random numbers.)
We calculate the mean-square displacements and the shear
relaxation moduli from the transient bond simulation data.

\begin{figure}[th]
 \begin{center}
 \includegraphics[width=.6\linewidth,clip]{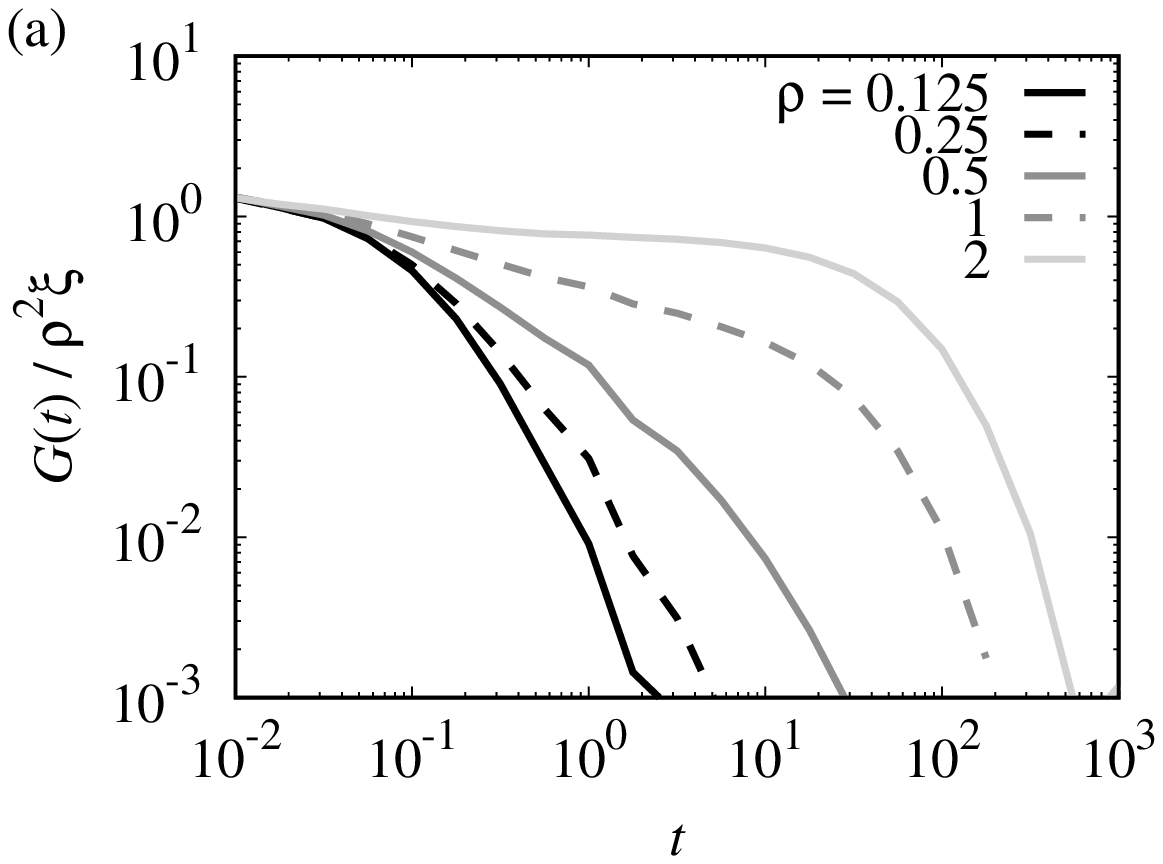}

 \includegraphics[width=.6\linewidth,clip]{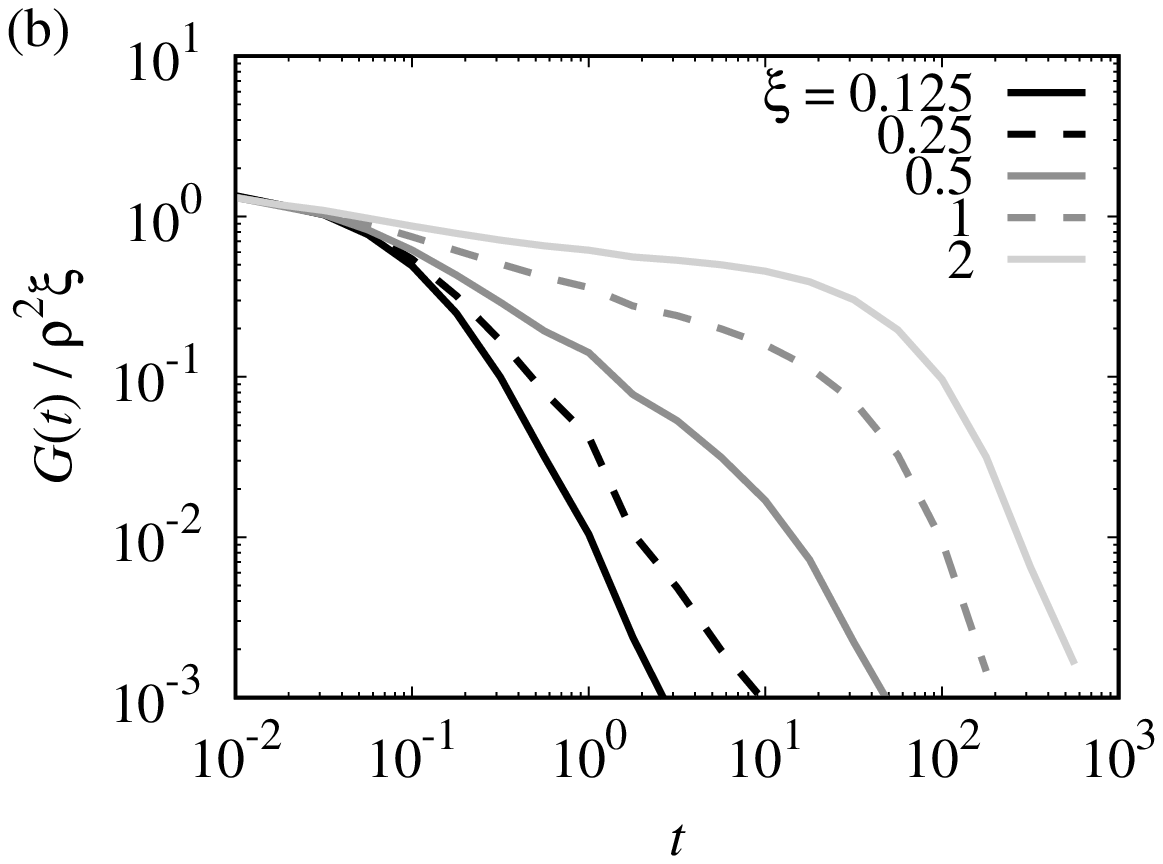}

 \includegraphics[width=.6\linewidth,clip]{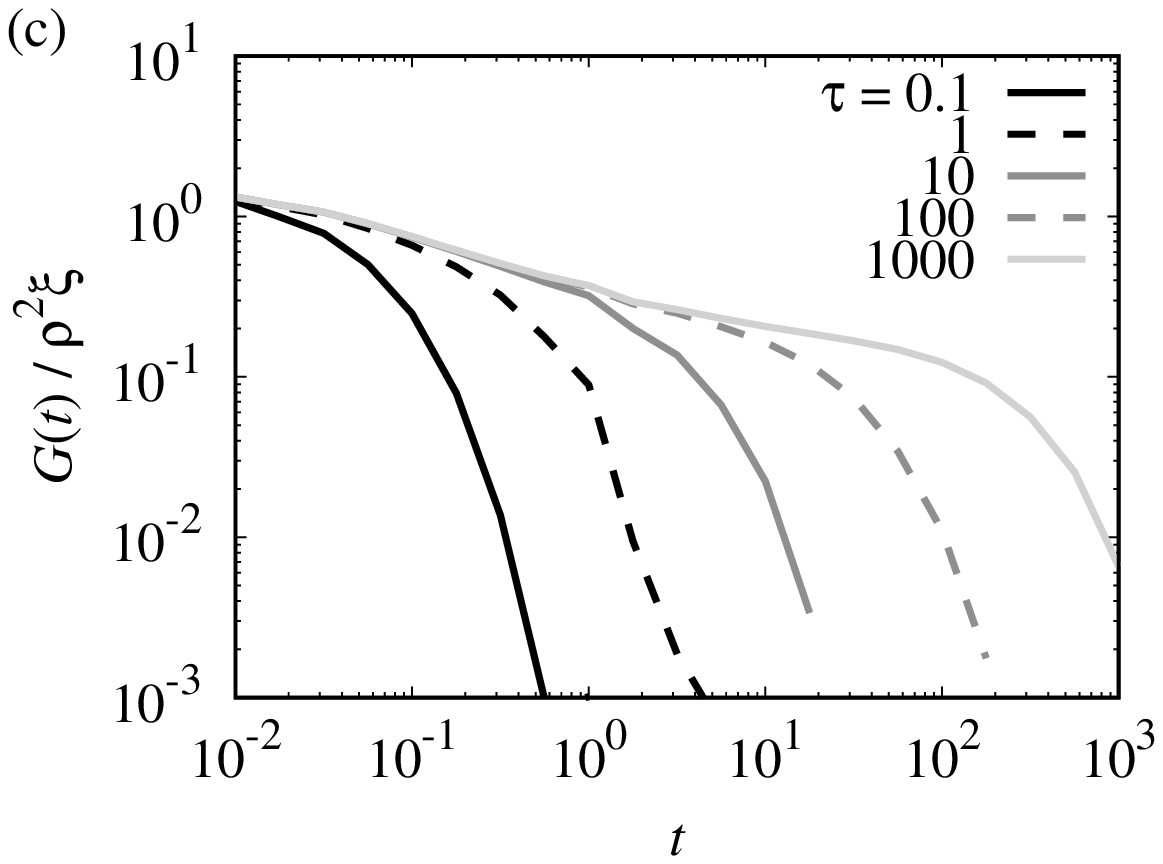}
 \end{center}
\caption{\label{relaxation_modulus_simulation_data}
 The relaxation modulus $G(t)$ by the ideal transient bond model with different
 values of (a) $\rho$, (b) $\xi$, and (c) $\tau$. The relaxation modulus
 is normalized by the factor $\rho^{2} \xi$ which is proportional to the
 bond density.}
\end{figure}

\begin{figure}[th]
 \begin{center}
 \includegraphics[width=.6\linewidth,clip]{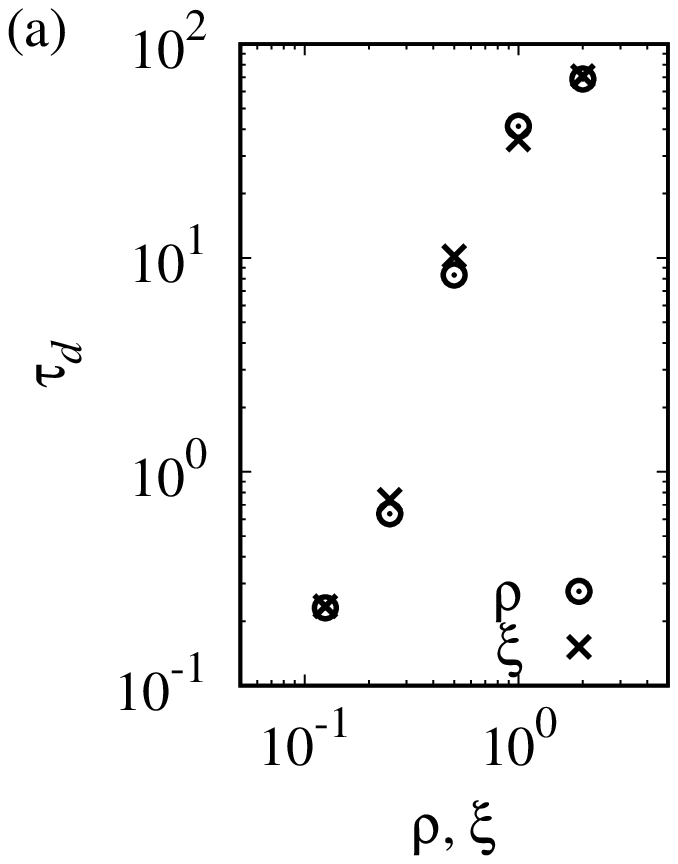}

 \includegraphics[width=.6\linewidth,clip]{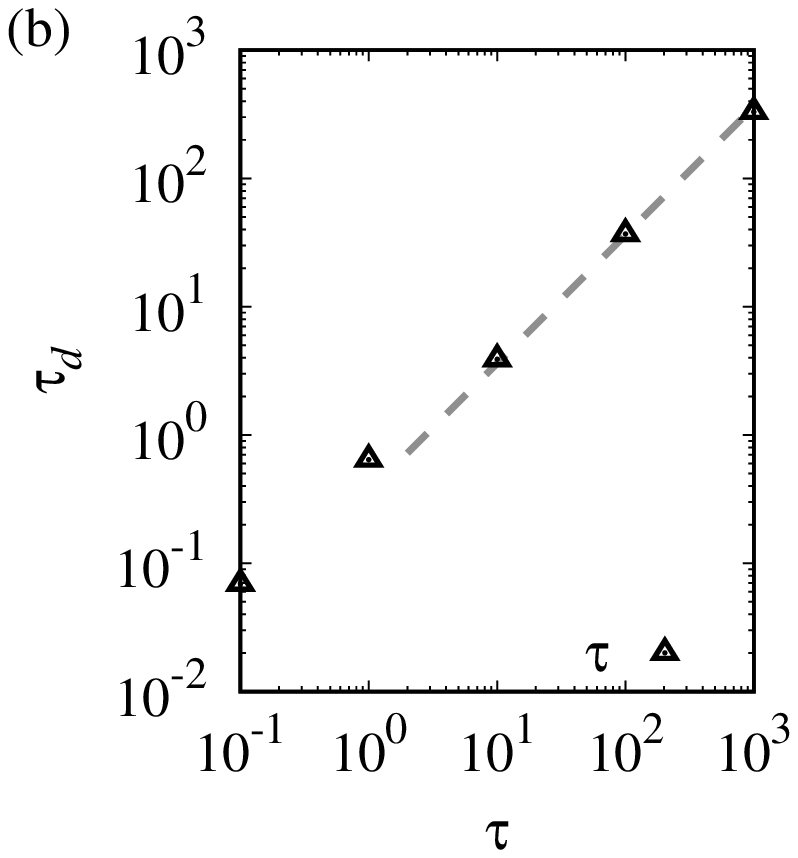}
 \end{center}
\caption{\label{relaxation_time_simulation_data}
 The longest relaxation time $\tau_{d}$ by the ideal transient bond
 model calculated from the relaxation modulus data. The dependence of
 $\tau_{d}$ on (a) $\rho$ and $\xi$, and (b) $\tau$. The dashed gray
 line shows the fitting result to the power-law type relation, $\tau_{d}
 \propto \tau^{1}$.}
\end{figure}

Fig.~\ref{relaxation_modulus_simulation_data} shows the relaxation
modulus data calculated from the simulation results with different
parameter sets. For convenience, the relaxation modulus is normalized by
$\rho^{2} \xi$ (which is proportional to the average transient bond
density $\phi${,} from eq~\eqref{average_bond_density}).
All the three parameters, $\rho$, $\xi$, and $\tau$
affect the relaxation modulus rather strongly. Especially, the
relaxation time increases as these parameters increase. The shapes of
the relaxation modulus {change} as the parameters change, and thus we
consider that the transient bond model can reproduce various viscoelastic
behavior by tuning the parameters.
The longest relaxation time $\tau_{d}$ is estimated from the relaxation
modulus at the long time region:
\begin{equation}
 \ln G(t) \approx \text{(const)} - \frac{\tau_{d}}{t} \quad (t \gtrsim
  \tau) .
\end{equation}
Fig.~\ref{relaxation_time_simulation_data} shows the dependence of the
longest relaxation time to parameters. From
Fig.~\ref{relaxation_time_simulation_data}(a), we observe that the
longest relaxation time strongly depends on $\rho$ and $\xi$. The
$\rho$-dependence of the longest relaxation time looks very similar to
the $\xi$-dependence. On the other hand, from
Fig.~\ref{relaxation_time_simulation_data}(b), we observe that the
effect of $\tau$ to the longest relaxation time is rather simple.
In the large $\tau$ region ($\tau \gtrsim 10$), the longest relaxation time is approximately proportional to the average
life time of the bond $\tau$ ($\tau_{d} \propto \tau$, see the dashed
line in Fig.~\ref{relaxation_time_simulation_data}(b)).
This result is physically natural because
in {our} model the stress relaxes when {a transient} bond is destroyed.

\begin{figure}[th]
 \begin{center}
 \includegraphics[width=.6\linewidth,clip]{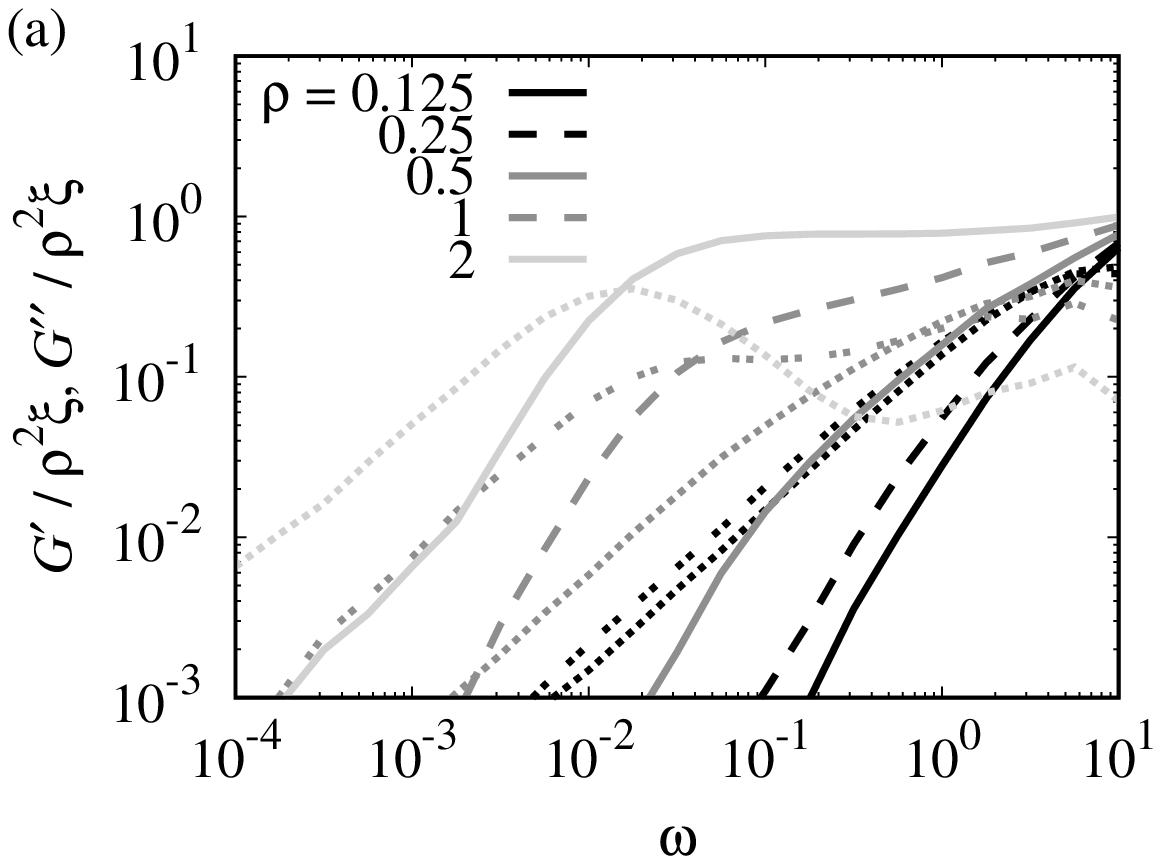}

 \includegraphics[width=.6\linewidth,clip]{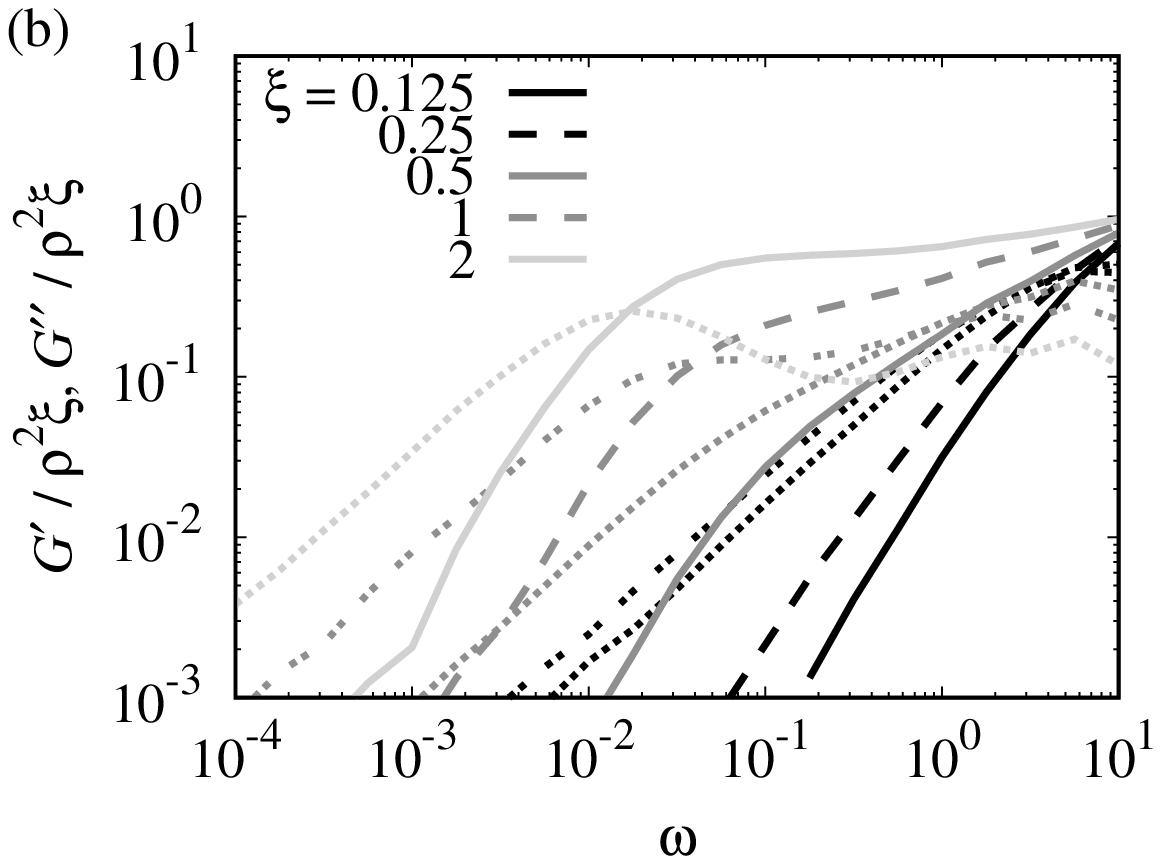}

 \includegraphics[width=.6\linewidth,clip]{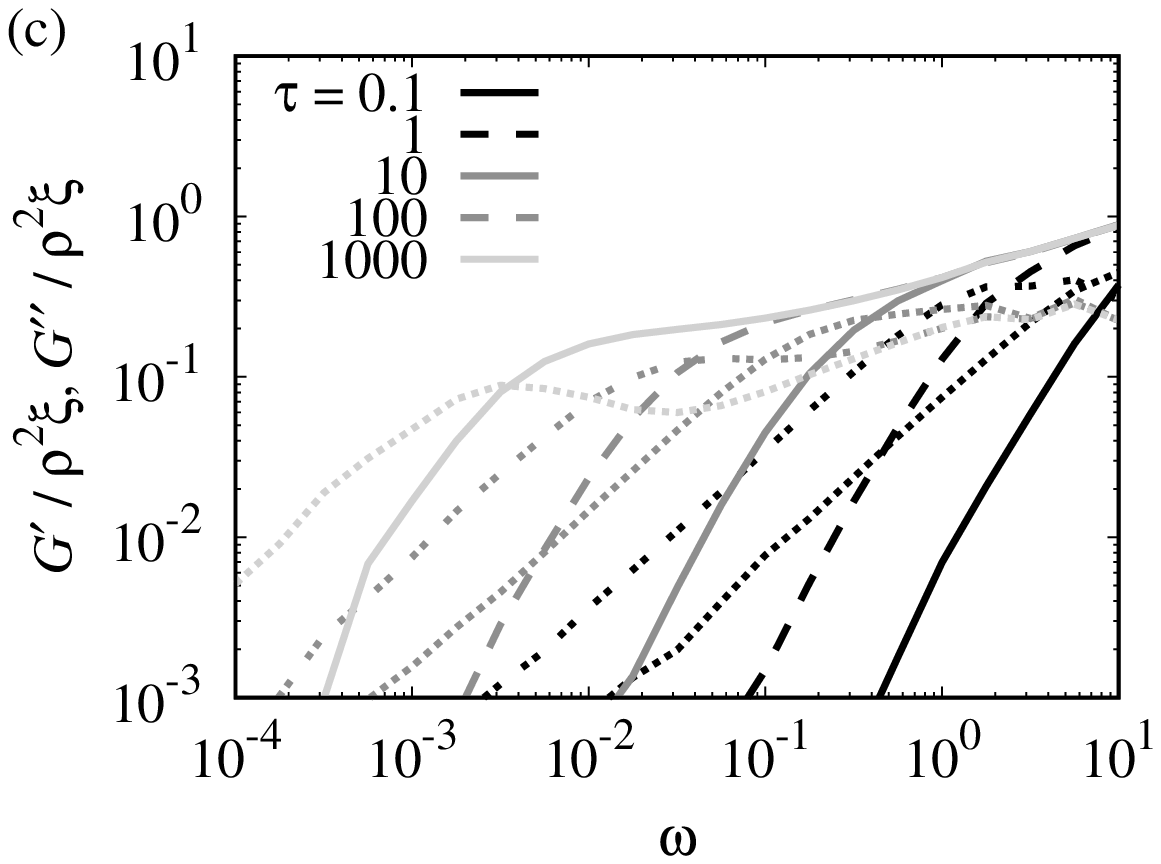}
 \end{center}
\caption{\label{storage_and_loss_moduli_simulation_data}
 The storage and loss moduli, $G'(\omega)$ and $G''(\omega)$ by the
 ideal transient bond model, calculated from the data in
 Fig.~\ref{relaxation_modulus_simulation_data}. The solid and dashed
 curves represent $G'$ and the dotted and dash-dotted curves represent $G''$.}
\end{figure}

In experiments, the storage and loss
moduli measured by {an} oscillatory {shear} mode are convenient and widely
utilized. We convert the relaxation modulus data into the storage and
loss moduli data, by performing the Fourier transform numerically:
\begin{align}
 \label{storage_modulus_from_relaxation_modulus}
 G'(\omega) & = \omega \int_{0}^{\infty} dt \, G(t) \sin (\omega t), \\ 
 \label{loss_modulus_from_relaxation_modulus}
 G''(\omega) & = \omega \int_{0}^{\infty} dt \, G(t) \cos (\omega t) .
\end{align}
Fig.~\ref{storage_and_loss_moduli_simulation_data} shows the storage and loss moduli for various parameter sets,
calculated from the relaxation modulus data in
Fig.~\ref{relaxation_modulus_simulation_data}.
As the relaxation modulus, the storage and loss moduli are normalized by 
 $\rho^{2} \xi$.
For the cases with large $\rho$ or $\xi$ values ($\rho = 2$ and $\xi =
2$ in Fig.~\ref{storage_and_loss_moduli_simulation_data} (a) and (b)),
the short time and long time relaxation modes become
separated. Other cases exhibit rather broad relaxation mode distributions.

\begin{figure}[th]
 \begin{center}
 \includegraphics[width=.6\linewidth,clip]{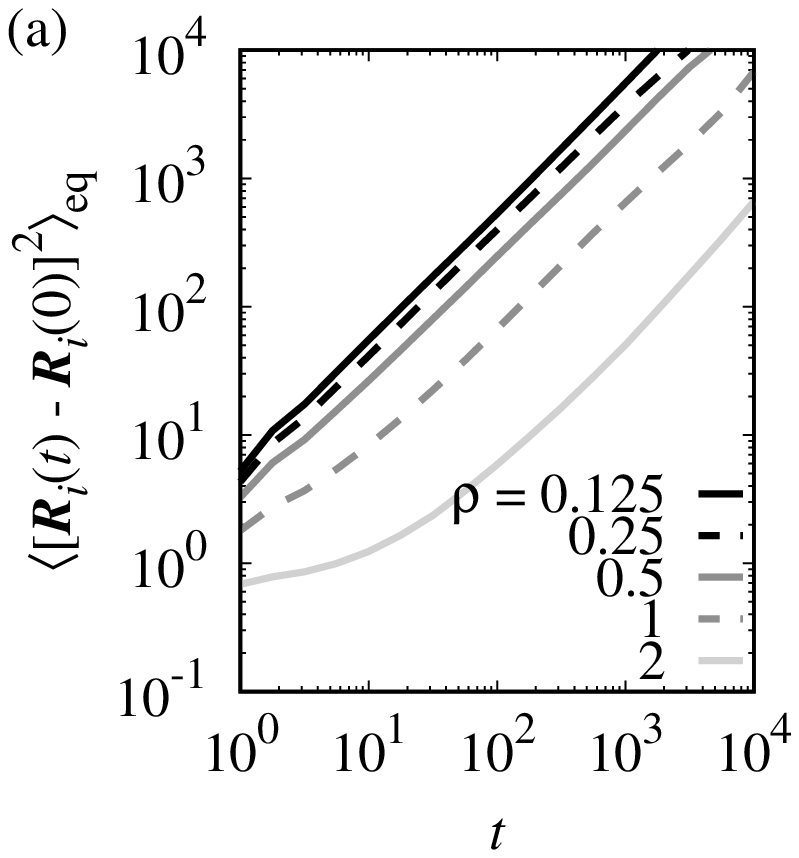}

 \includegraphics[width=.6\linewidth,clip]{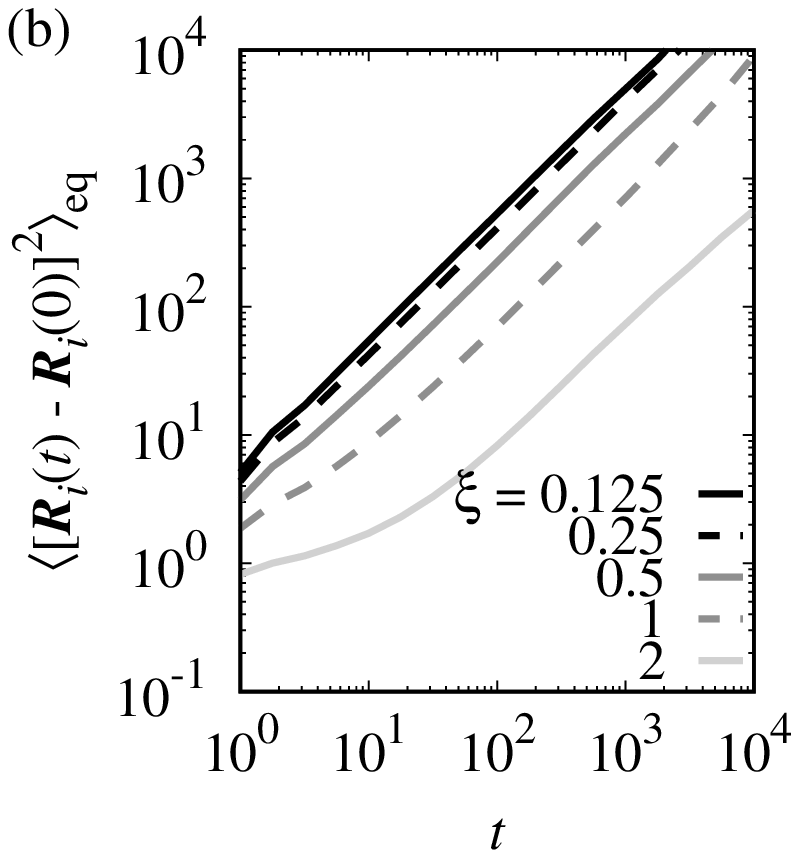}

 \includegraphics[width=.6\linewidth,clip]{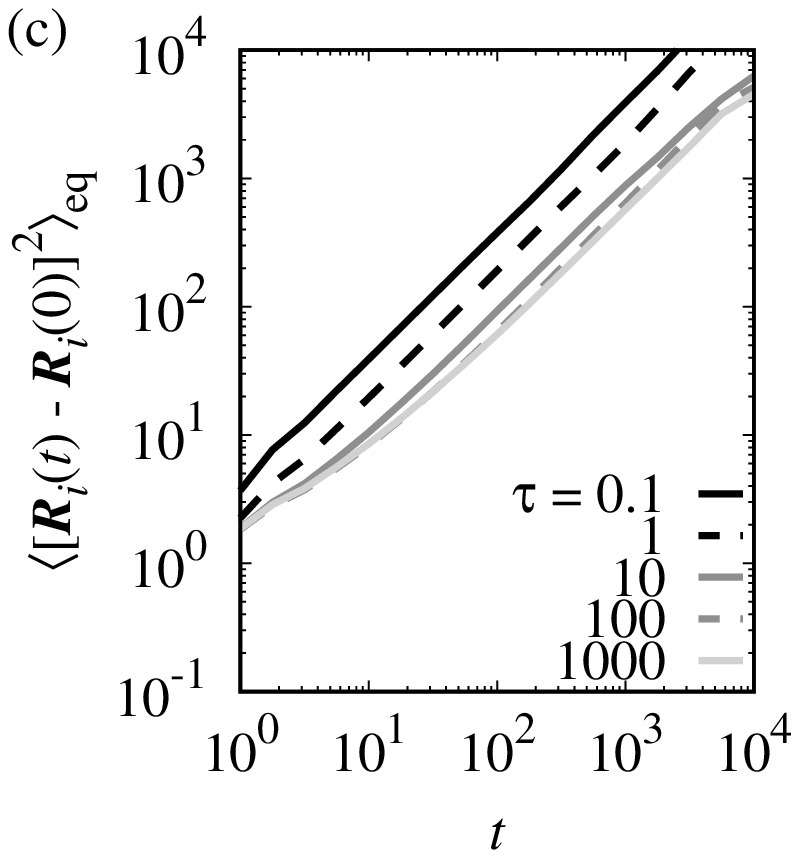}
 \end{center}
\caption{\label{msd_simulation_data}
 The mean-square displacement by the ideal transient bond model with
 different values of (a) $\rho$, (b) $\xi$, and (c) $\tau$.}
\end{figure}

\begin{figure}[th]
 \begin{center}
 \includegraphics[width=.6\linewidth,clip]{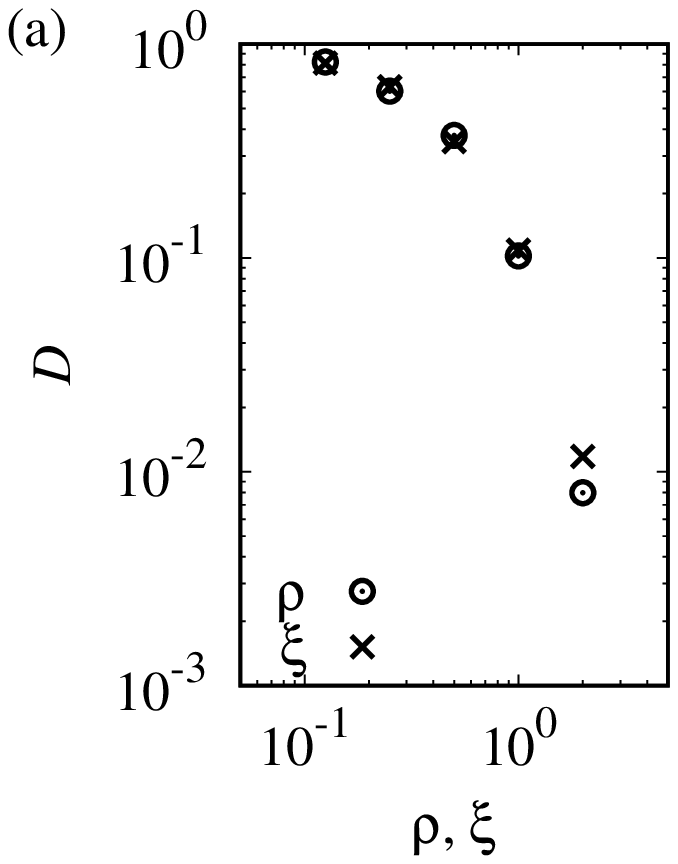}

 \includegraphics[width=.6\linewidth,clip]{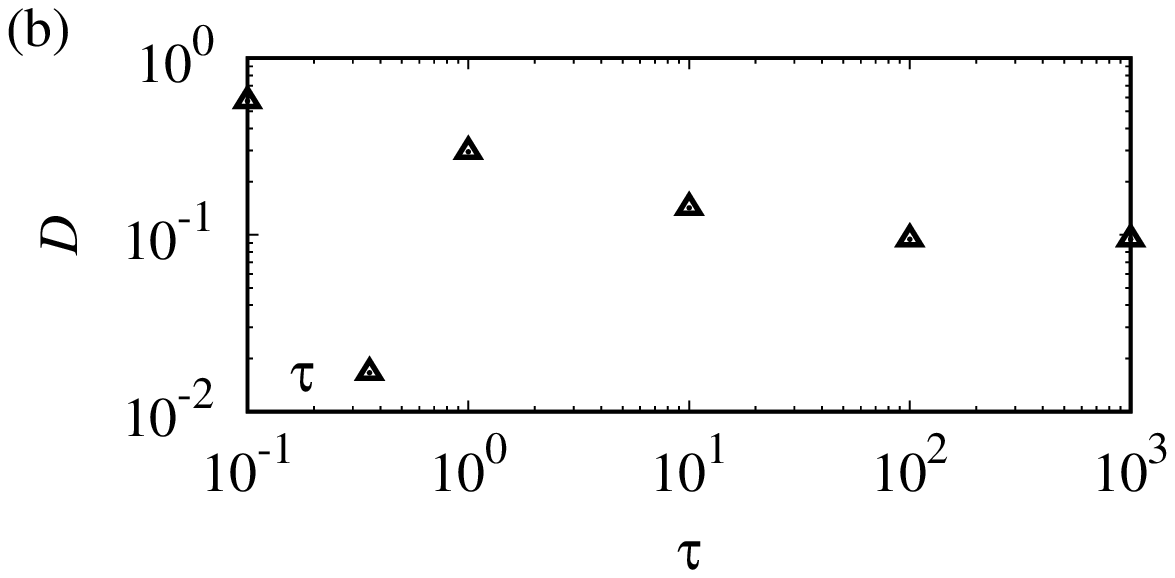}
 \end{center}
\caption{\label{diffusion_coefficient_simulation_data}
 The diffusion coefficient $D$ by the ideal transient bond model, calculated
 from the mean-square displacement data in
 Fig.~\ref{msd_simulation_data}.
 The dependence of the diffusion coefficient $D$ on (a) $\rho$ and
 $\xi$, and (b) $\tau$.}
\end{figure}

Fig.~\ref{msd_simulation_data} shows the dependence of the mean-square
displacements to parameters $\rho$, $\xi$, and $\tau$. The mean-square displacement decreases as 
one of the three parameters, $\rho$, $\xi$, and $\tau$, increases. This
is consistent with the results for the relaxation modulus. 
The mean-square displacements exhibit the subdiffusion behavior for
relative short time regions. The diffusion coefficient $D$ is estimated
from the mean-square displacement at the long time region:
\begin{equation}
\ln \langle [\bm{R}_{i}(t) -
  \bm{R}_{i}(0)]^{2} \rangle \approx \ln (6 D) + \ln t \quad (t \gtrsim \tau) .
\end{equation}
If there is no transient bonds in the system (the ideal Brownian gas),
the diffusion coefficient becomes $D = 1$ in the dimensionless unit.
Fig.~\ref{diffusion_coefficient_simulation_data} shows the diffusion
coefficients estimated from the mean-square displacement data in
Fig.~\ref{msd_simulation_data}. {For the cases where} the parameters $\rho$, $\xi$,
{or} $\tau$ {is} small, the diffusion coefficient approaches to $D = 1$, as
expected.
As shown in Fig.~\ref{diffusion_coefficient_simulation_data}(a),
the $\rho$- and $\xi$-dependence of
the diffusion coefficient is similar, as the case of the relaxation time.
The $\tau$-dependence of the diffusion coefficient 
{is weak}. 
The data in Fig.~\ref{diffusion_coefficient_simulation_data}(b) can be
fitted to a power-law type relation, $D \propto \tau^{-0.27}$, for
the relatively small $\tau$ region, and to constant for the relatively
large $\tau$ region. This result seems not to be consistent with the
naive expectation from the relaxation time data. We will discuss the
diffusion mechanism in the next section.

From {these} simulation data, we can {roughly} summarize the behavior of
the ideal transient bond model as follows. All the three parameters ($\rho$,
$\xi$, and $\tau$) strongly
affect the {viscoelastic} relaxation data, whereas the diffusion data are
strongly affected by $\rho$ and $\xi$, and weakly affected by $\tau$.

\subsection{Entangled Polymer Systems}

We can map the entangled polymers to the ideal transient bond model,
when we employ a reference parameter set and use the relations in Sec.~\ref{parameters_for_entangled_polymer_systems}.
Here we employ $\rho = 1$, $\xi = 1$, and $\tau = 100$
as the reference parameter set for $N = 1$. Other simulation conditions
are the same as Sec.~\ref{effect_of_dimensionless_parameters}.

\begin{figure}[th]
 \begin{center}
 \includegraphics[width=.6\linewidth,clip]{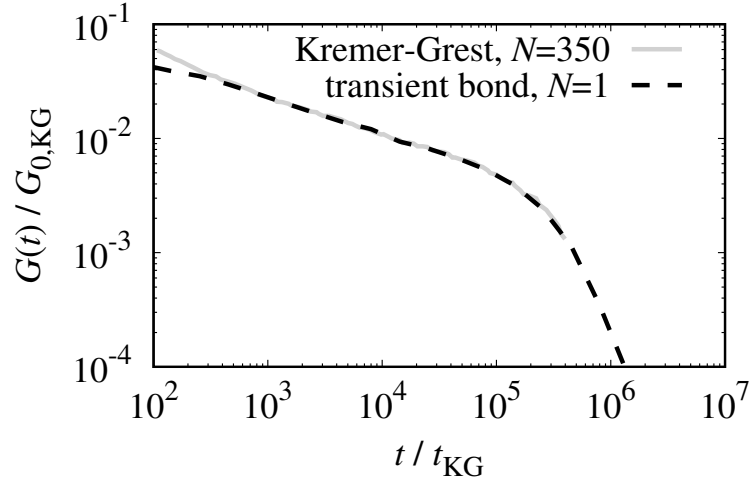}
 \end{center}
\caption{\label{fitting_to_kremer_grest}
 Comparison of the relaxation modulus data by the Kremer-Grest model ($N
 = 350$) \cite{Likhtman-Sukumaran-Ramirez-2007} and
 the ideal transient bond model ($N = 1$). The time and modulus are normalized by
 the unit time scale and the unit modulus (stress) scale of the
 Kremer-Grest model, $t_{\text{KG}}$ and $G_{0,\text{KG}}$.}
\end{figure}

The shear relaxation modulus
and the storage and loss moduli
for the reference parameter set
(in Fig.~\ref{relaxation_modulus_simulation_data} and
Fig.~\ref{storage_and_loss_moduli_simulation_data})
look similar to those of mildly entangled polymers
obtained by various mesoscopic simulation models and experiments. In
fact, by rescaling the time and stress scales, we can map the storage
and loss moduli to those obtained by the Kremer-Grest
model\cite{Likhtman-Sukumaran-Ramirez-2007} as shown
in Fig.~\ref{fitting_to_kremer_grest}.
The degree of polymerization (number of beads per chain) in the Kremer-Grest model is
$N_{\text{KG}} = 350$, thus we have the scale conversion factor
$N_{\text{TB}} / N_{\text{KG}} = 3.5 \times 10^{2}$, where
$N_{\text{TB}}$ is the degree of polymerization in the transient bond model.
The scale conversion factors for the time and modulus (stress) can be
determined as the rescaling {(or shift)} factors used in
Fig.~\ref{fitting_to_kremer_grest}.
The scale conversion factor for the time scales of the transient bond
model $t_{\text{TB}}$ and the Kremer-Grest model $t_{\text{KG}}$ is $t_{\text{TB}} / t_{\text{KG}} = 8.0
\times 10^{3}$, and one for the modulus (stress) scales of the transient
bond model $G_{0,\text{TB}}$ and the Kremer-Grest model
$G_{0,\text{KG}}$ is $G_{0,\text{TB}} / G_{0,\text{KG}} = 3.3 \times
10^{-2}$.

\begin{figure}[th]
 \begin{center}
 \includegraphics[width=.6\linewidth,clip]{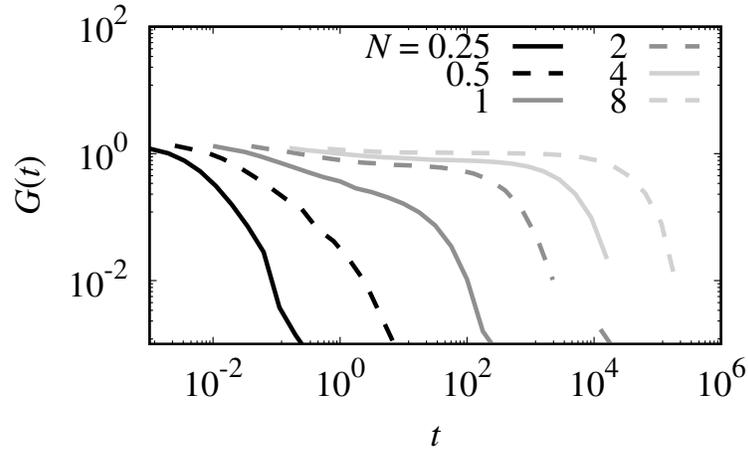}
 \end{center}
\caption{\label{relaxation_modulus_reptation}
 The relaxation modulus $G(t)$ of entangled polymer systems with
 different degrees of polymerization $N$.}
\end{figure}

\begin{figure}[th]
 \begin{center}
 \includegraphics[width=.6\linewidth,clip]{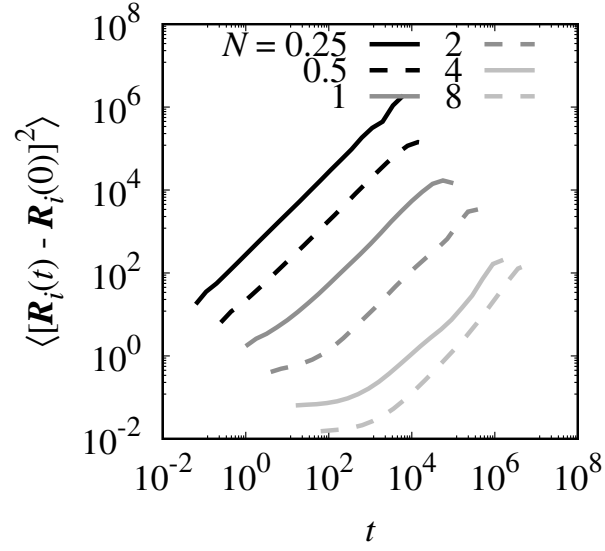}
 \end{center}
\caption{\label{msd_reptation}
 The mean-square displacement data of entangled polymer systems with
 different degrees of polymerization $N$.}
\end{figure}

\begin{figure}[th]
 \begin{center}
 \hspace{-0.1\linewidth}\includegraphics[width=.6\linewidth,clip]{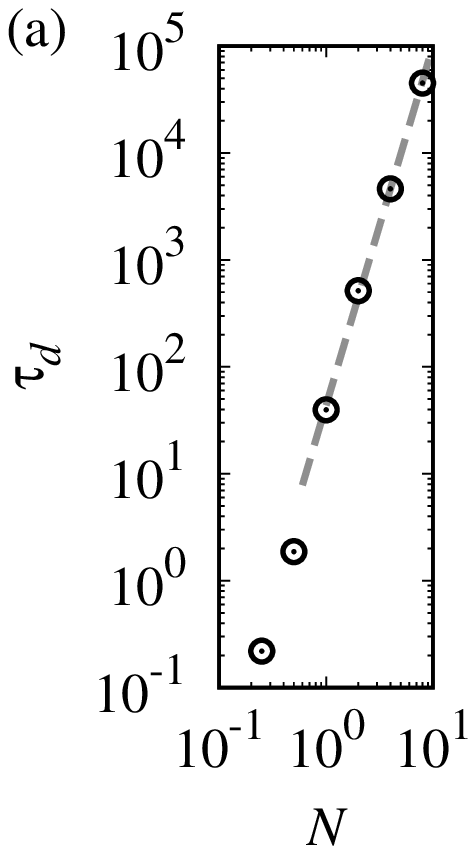}
 \hspace{-0.3\linewidth}\includegraphics[width=.6\linewidth,clip]{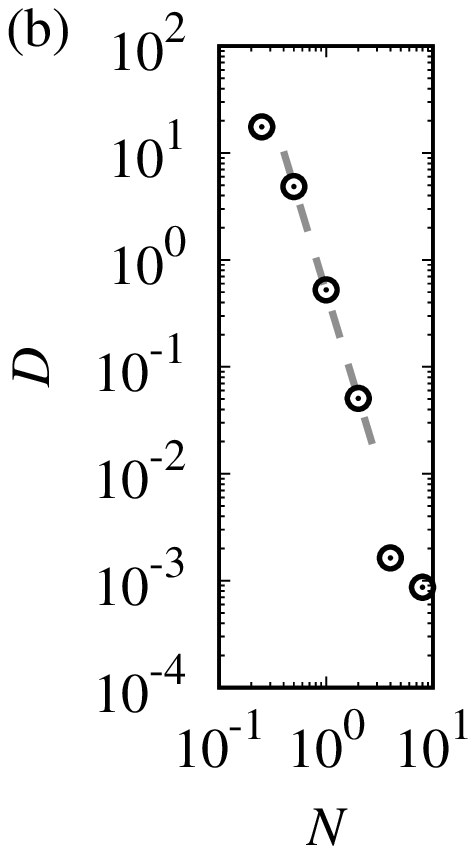}
 \end{center}
\caption{\label{relaxation_time_diffusion_coefficient_reptation}
 (a) The longest relaxation time $\tau_{d}$ of the entangled polymers
 calculated by the relaxation modulus shown in
 Fig.~\ref{relaxation_modulus_reptation}. 
 (b) The diffusion coefficient $D$ of the entangled polymers. The gray
 dashed lines show the fitting results to the power-law type relations,
 $\tau_{d} \propto N^{3.4}$ and $D \propto N^{-3.3}$.}
\end{figure}

We show the shear relaxation modulus data for other values of $N$
($N = 0.25, 0.5, 2, 4$, and $8$) in Fig.~\ref{relaxation_modulus_reptation}. The relaxation modulus for
large $N$ seems to be sharp. {Naively, we consider that} this would be {because of} the lack of the
short-time relaxation modes due to the $N$-dependence of the coarse-graining
level.
Fig.~\ref{msd_reptation} shows the mean-square displacement data for the same
values of $N$ as in Fig.~\ref{relaxation_modulus_reptation}. The
mean-square displacements shown in Fig.~\ref{msd_reptation} correspond
to the mean-square displacements of the centers of mass of polymer chains
(so-called $g_{3}(t)$ \cite{Kremer-Grest-1990}). The mean-square
displacement of the center of mass is known to
exhibit the crossover behavior \cite{Kremer-Grest-1990}:
\begin{equation}
 \label{msd_crossover}
 \langle [\bm{R}_{i}(t) -
  \bm{R}_{i}(0)]^{2} \rangle \propto
  \begin{cases}
   t^{1} & (t \lesssim \tau_{e}) , \\
   t^{1/2} & (\tau_{e} \lesssim t \lesssim \tau_{R}) , \\
   t^{1} & (\tau_{R} \lesssim t) ,
  \end{cases}
\end{equation}
where $\tau_{e} \propto N^{0}$ and $\tau_{R} \propto N^{2}$ are the entanglement time and the Rouse
time, respectively\cite{Doi-Edwards-book}. The data shown in Fig.~\ref{msd_reptation} are not
consistent with eq \eqref{msd_crossover}. (For large $N$ cases,
we can observe the crossover of the mean-square displacement from
constant to the normal diffusion.)
To quantitatively analyze the $N$-dependence of the
viscoelastic and diffusion behavior, we calculate the longest relaxation
time $\tau_{d}$ and the diffusion coefficient $D$.
The longest relaxation time $\tau_{d}$ and the diffusion
coefficient $D$ of entangled polymer systems are shown in
Fig.~{\ref{relaxation_time_diffusion_coefficient_reptation}}. The
longest relaxation time data for relatively large $N$ can be fitted to
the power-law, $\tau_{d} \propto N^{3.4}$. This power-law exponent is
consistent with the experimental data, the theoretical prediction, and the simulation data by various
simulation models\cite{Doi-Edwards-book}. The diffusion coefficient can be fitted to the
power-law, $D \propto N^{-3.3}$, but this power-law exponent is not
consistent with experimental data and other simulation models\cite{Lodge-1999}.

From these simulation results,
we conclude that the viscoelastic behavior of the entangled polymers can be reasonably
reproduced if the parameters are determined based on the reference
parameter set and the degrees of polymerization.
However, the diffusion behavior is not reproduced well.
We discuss the effect of the parameters and the mapping {of
the transient bond model} to the entangled polymer
systems in detail, in the next section.

\section{Discussions}
\label{discussions}

{\subsection{Relaxation and Diffusion Behavior of Transient
  Bond Model}}

As shown in Figs.~\ref{relaxation_modulus_simulation_data}-\ref{storage_and_loss_moduli_simulation_data}{,}
the relaxation and diffusion data exhibit similar $\rho$- and $\xi$-dependence.
This can be intuitively understood if we consider
the average number of transient bonds per particle. From eq
\eqref{average_bond_density}, the average number of transient bonds per
particle is estimated as
\begin{equation}
 \frac{\phi}{\rho} = \frac{\xi \rho}{2}
  \left(\frac{2 \pi a^{3}}{3}\right)^{3/2} \propto \xi \rho .
\end{equation}
Under the mean-field approximation (which is shown in Appendix~\ref{mean_field_approximation_for_single_particle_dynamics}), we expect that the relaxation behavior of a single
particle in the system is determined solely by the average number of
transient bonds attached to {a} target particle. Then, the relaxation
behavior is approximately determined by the factor $\xi \rho$, and thus the
relaxation behavior will be similar if $\xi \rho$ is the same.
Conversely, we can {almost fully} tune the relaxation behavior of the
system only by two parameters, $\xi$ and $\tau$,
even if we fix $\rho$ to be constant. Such a reduction of the number of free
parameters will be useful when we fit the model parameters to
{a} specific target system.

If the value of $\rho \xi$ is sufficiently large, one particle in the
system will be strongly constrained by many transient bonds. The
relaxation occurs only through the bond reconstruction
process{, and}
the characteristic time {of the reconstruction is} $\tau$. Based on this picture, we expect that
the longest relaxation time will approach to $\tau$ for large $\rho \xi$
cases. In Fig.~\ref{relaxation_time_simulation_data}(a), we can
observe that the longest relaxation time actually approaches to $\tau =
100$ {for large $\rho$ or large $\xi$ cases}. In addition, we expect
that the relaxation
{function}
will approach to a single exponential form with the relaxation time
$\tau$. In other words, the fluctuation of the number of transient bonds
attached to one particle will broaden the relaxation mode distribution.
Therefore, roughly speaking, we can utilize $\rho$ or $\xi$ to tune the shape of the
relaxation mode distribution and utilize $\tau$ to tune the longest
relaxation time. These properties are consistent with the estimates
under the mean-field approximation in Appendix~\ref{mean_field_approximation_for_single_particle_dynamics}.

If the value of $\rho \xi$ is sufficiently small, we expect that the
transient bonds {cannot form network structures and they}
can form only dumbbell-like structures (dimers). The
longest relaxation time is estimated as one of {a} dimer ({a} dumbbell
with a harmonic spring\cite{Kroger-2004}), and thus we
have $\tau_{d} \approx 1 / 12$. It should be noted that
this relaxation time of {a dimer} is independent of the bond life time
$\tau$. In addition, the contribution of dimers to diffusion coefficient is considered
to be small, and thus the diffusion coefficient becomes $D \approx 1$ as
we mentioned. This is again independent of the bond life time.
Thus the dynamical behavior will be almost independent of the
bond life time $\tau$, if $\rho \xi$ is sufficiently small.

The effects of the average life time of the bond $\tau$ on the relaxation
time and the diffusion coefficient {apparently} seem not to be {consistent}. Unlike the
simple $\tau$-dependence of the longest relaxation time, the $\tau$-dependence
of the diffusion coefficient seems to be very weak. Moreover, in the large $\tau$
region ($\tau \gtrsim 100$), the diffusion coefficient becomes almost independent of $\tau$.
This can be understood as follows. A particle in the system takes two
states; the free state in which no bonds are attached to the particle,
and the constrained state in which bonds are attached to the
particle. In the free state, the diffusion of the particle is not
constrained and thus we will observe the free diffusion. In the
constrained state, the particle is constrained by the bonds and the
average position is almost fixed. The particle at the constrained state
can diffuse {only} via the reconstruction of bonds, thus the diffusion
coefficient at the constrained state is inversely proportional to the
life time. The average diffusion coefficient is the average of the
diffusion coefficients at these {two} states. For sufficiently large $\tau$,
the diffusion coefficient at the constrained state becomes negligibly
small and thus we observe the $\tau$-independent diffusion
coefficient. (See
Appendix~\ref{mean_field_approximation_for_single_particle_dynamics} for
detailed calculations.) This mechanism is somewhat similar to the dynamic
heterogeneity observed in supercooled or glassy systems\cite{Yamamoto-Onuki-1998,Sillescu-1999,Uneyama-Miyaguchi-Akimoto-2015}.
The transient bond model may be utilized as a model for supercooled or
glassy systems which exhibit dynamic heterogeneity.

{\subsection{Entangled Polymer Systems}}

The simulations for the entangled polymer systems by the ideal transient bond
model showed reasonable results for the viscoelastic data. The
dependence of the relaxation time $\tau_{d}$ to the degree of
polymerization $N$ is consistent with the well-known power law,
$\tau_{d} \propto N^{3.4}$\cite{Doi-Edwards-book}.
This result is rather surprising because there is 
no contour length fluctuation (CLF) in the transient bond model. The
pure reptation model gives the power-law exponent $3$, and
the exponent $3.4$ is believed to be the apparent exponent due to the correction by the
CLF \cite{Doi-Edwards-book}. Our result implies that the CLF may not be essential to reproduce
the exponent $3.4$. We expect that the fluctuation of the positions of
particles may give the correction to the relaxation modulus.
Some experiments\cite{Liu-Halasa-Keunings-Bailly-2006,Matsumiya-Kumazawa-Nagao-Urakawa-Watanabe-2013} and simulations\cite{Masubuchi-Amamoto-Pandey-Liu-2017}
report that the CLF mechanism is affected by the constraint-release
(CR), and in absence of the CR mechanism, the power-law exponent
apparently becomes lower than $3.4$. In the transient bond model, the
contribution of the CR type mechanism clearly depends on $N$, and thus
the CR might affect the relaxation behavior in a different way from the
reptation model. This might be one possible origin of the exponent
$3.4$.

The fact that the shear relaxation modulus of the transient bond can be
fitted well {to that} by the Kremer-Grest model is also surprising.
The transient bond model is highly coarse-grained, and is designed to
reproduce the dynamics at the long time region. Nevertheless, the shear
relaxation modulus by the transient bond model agrees well with {that} by
the Kremer-Grest model even at the short time, Rouse relaxation region.
{Of course, this apparent Rouse like relaxation behavior
may be just an artifact. In Fig.~\ref{relaxation_modulus_reptation},
we cannot find such Rouse like behavior for
systems with large $N$. As we discussed, the relaxation modulus is
expected to approach the single exponential form for sufficiently large
$N$ because $\rho \xi \propto N$. Then we will only have well-developped
plateau at the short time region. Even if the Rouse like behavior is
just a model artifact,}
the reason why we have such a {power-law type} behavior is
not clear.
One possible mechanism is that the modulation of the relaxation mode distribution due
to the formation of {a network-structure}. The power-law like viscoelastic
behavior at the short time region was experimentally observed in
network-forming associative telechelic polymer solutions
\cite{Uneyama-Suzuki-Watanabe-2012}.
{It is plauseble that} the spatial coupling of transient
bonds will give similar power-law like relaxation behavior at the short time
scale, which (accidentally) has the same power-law exponent as the Rouse model.

In contrast to the viscoelastic behavior, the dependence of the diffusion
coefficient $D$ to the degree of polymerization $N$ is not consistent with the well-known power law
$D \propto N^{-2.4}$\cite{Lodge-1999}.
We consider that this is due to the lack of the reptation motion in the
transient bond model. As we discussed, the diffusion
coefficient is determined as the average of the diffusion coefficients
at the free and constrained states. The fraction of {the} free state decreases
as $N$ increases.
Also, the diffusion coefficients at the
constraint states strongly decrease as $N$ increases. Thus we have very
strong $N$-dependence of the diffusion coefficient.

We will need to incorporate
some mechanisms which reproduce the reptation like diffusion motion to
the model, to recover the diffusion behavior which is consistent with
experiments and other simulation models. 
The introduction of another dynamics rule to the model which enhances the
diffusion will improve the diffusion behavior.
One possible way is to move particles
without changing bonds. Such a motion can be realized, for example, if we
stochastically exchange the positions of two bonded particles.
{Other} possible {ways are} to generalize the model and include the
conformational degrees of freedom\cite{Fitzgerald-Briels-2018}, and to
introduce the reptation type diffusion dynamics {by
modifying the dynamic equations}. Because
we have the explicit expression of the effective free energy, such
extensions of the model will be rather straightforward.

Although the diffusion behavior of entangled polymers cannot be reproduced well by the transient
bond model {(at least in the current form)}, the viscoelastic behavior can be reasonably
reproduced. Therefore, we expect that the transient bond model can be
utilized to simulate viscoelasticity. Because the transient bond model is highly
coarse-grained and is computationally efficient, it will be useful when
we are interested only on the viscoelasticity.
The relation among our model and other mesoscopic models for entangled
polymer systems is interesting. By determining the scale conversion
factors among different models, we can connect or compare the simulation
data by different models\cite{Masubuchi-Uneyama-2018}.
The detailed comparison among our model and some mesoscopic coarse-grained models for entangled
polymers is in progress and will be published in near future.

\

{\subsection{Possible Extensions of Model}}

In this work we performed simulations for the simplest, ideal case of the
transient bond model{, to investigate the basic properties
of the model. We set the interaction potential
between particles, $v(\bm{r})$, to be zero, but this is not realistic for
polymer melts. The compressiblity of a polymer melt is generally very
low, whereas one of the ideal transient bond model is
rather high. In the original RaPiD model, Kindt and Briels\cite{Kindt-Briels-2007} employed the
Gaussian repulsive potential to repel particles.
We will also need to employ
the Gaussian repulsive potential, to perform more realistic simulations
for entangled polymers. However, the entanglement effect
exists even at the limit of the zero excluded volume, as far as the
chains cannot cross each other. Thus
we expect that the ideal transient bond model can capture the
characteristic dynamical behavior of entangled polymers even in absence
of the Gaussian repulsive potential like the original RaPiD model.
It would be worth mentioning here that the
multi-chain slip-spring model\cite{Uneyama-Masubuchi-2012} can reproduce
dynamical properties of entangled polymers even without any interaction
potential between segments.}

{Although in this work we limit ourselves to the ideal systems,}
the transient bond model can be applied to much
complicated systems by using different potentials {and
dynamics models. Instead of the Langevin equation for the particles,
other dynamic equation models can be employed.}
For example, if we use the dynamic
equation {of the dissipative particle
dynamics (DPD) model}\cite{Espanol-Warren-1995,Kinjo-Hyodo-2007}, which conserves the
momentum, we will be able to simulate the complex flow of
viscoelastic materials. Langeloth and coworkers
\cite{Langeloth-Masubuchi-Bohm-MullerPlathe-2013} showed that the combination of the multi-chain slip-spring
model and the DPD dynamic equation reproduces dynamic properties of
entangled polymer melts and solutions reasonably. Due to its high
coarse-graining level, the combination of the
transient bond model and the DPD dynamic equation will be
computationally more efficient than the mutli-chain slip-spring model.
The reconstruction dynamics of the transient bonds can be also
changed. {In some systems, the destruction rate may depend
on the bond number or the bond vector.}
The destruction and construction rates can be changed as
long as they satisfy the detailed balance condition. Because our model
is based on the effective free energy, such a modification is rather
straightforward.
{The applications of our model to star polymers and
telechelic associative polymers will be intersting.}

{The incorporation of the transient potential
\cite{Kindt-Briels-2007,SantosdeOliveira-Fitzgerald-denOtter-Briels-2014,Sprakel-Spruijt-vanderGucht-Padding-Briels-2009,Sprakel-Padding-Briels-2011,
Liu-Padding-denOtter-Briels-2013,Liu-denOtter-Briels-2014,Fitzgerald-Lentzakis-Sakellariou-Vlassopoulos-Briels-2014,Fitzgerald-Briels-2018}
is powerful and promising method to model soft matter systems which
exhibit complex dynamical behavior.
We expect that our approach to unify the RaPiD and slip-spring model
can be further generalized, by adding extra degrees of freedom to the
{system}. For example, we can add the average life times of bonds as the
extra degrees of freedom, in a similar way to the slip-link model\cite{Khaliullin-Schieber-2009}.
A recent work\cite{Uneyama-Miyaguchi-Akimoto-2015} showed that
the diffusion coefficient (or the friction coefficient) should be
treated as a fluctuating variable in some systems.
In such systems, the diffusion
coefficient would be interpreted as extra degrees of freedom to modulate the dynamics, just
like the transient bonds in our model. Modeling with extra degrees of
freedom to modulate dynamics will be useful for various systems, as an
alternative way to the generalized Langevin equation with a memory kernel\cite{Kawasaki-1973}.
Our approach would be informative
to construct and analyze these dynamical models.}

\

\section{Conclusion}

We constructed the transient bond model, based on the {ideas} of the RaPiD
model and the multi-chain slip-spring model. Our transient bond model
has the {well-defined} effective free energy, and the transient bonds affect only
dynamical properties. As the simplest, ideal case, we considered the
ideal transient bond model in which the equilibrium properties reduce to
those of {an} ideal gas. The ideal transient bond model can be
characterized by three dimensionless parameters, the particle density ${a^{3}}\rho$, the
effective fugacity $\xi$, and the average life time of transient bonds
$\tau{/\tau_{\text{TB}}}$. The effects of these parameters to the linear viscoelasticity and
the mean-square displacement were investigated by simulations. The
parameters $\rho$ and $\xi$ have similar effects on {the dynamical
quantities}, because the dynamical behavior is determined by the average
number of transient bonds per {particle}. For entangled polymer systems, we
derived the relation between the degree of polymerization and the
parameters in the ideal transient bond model. The linear viscoelasticity of the
entangled polymer systems can be reasonably reproduced by the ideal
transient bond model. However, the diffusion behavior cannot be reproduced by the
ideal transient bond model. Thus the transient bond model can be
utilized to study entangled polymer systems when we are interested only
in the viscoelastic behavior. The transient bond model has a simple
structure {and it} can be tuned {for a} specific
{traget system}. Also, it can be
combined with other mesoscopic models. The application and extension of
the transient bond model to much complicated systems will be the future work.

\section*{Acknowledgment}

The author thanks Prof.~Yuichi Masubuchi {and Prof.~Wim Briels} for helpful comments.
This work was supported by Grant-in-Aid (KAKENHI) for Scientific
Research C 16K05513 and Grant-in-Aid (KAKENHI) for Scientific
Research A 17H01152.

%------------------------------------------------------------------------------
\appendix

\section{Approximation for Large Bond Number}
\label{approximation_for_large_bond_number}

In this appendix, we consider the relation between our transient bond
model shown in the main text and the {(original)} RaPiD model\cite{Kindt-Briels-2007}. In the RaPiD model,
the dynamics of the system is described by two Langevin
equations. One is the Langevin equation for the position, and is almost
the same as that in our model{, and} another is the Langevin equation for the
bond number (in the RaPiD model, the bond number is treated as
{a} continuum variable). The dynamics for the bond number in our model is
the birth-death type dynamics, and two models apparently seem to be different.

We consider the case where the number of bonds between the $i$-th and
$j$-th particles is large. We consider the dynamics of the bond number
$n_{ij}$ under the condition where other variables are fixed.
For simplicity, we describe the number of
transient bonds as $n$. We express the probability distribution of the bond number $n$
as $P(n,t)$. Then, the dynamic equation for $P(n,t)$ can be expressed as
the following master equation:
\begin{equation}
 \label{master_equation_bond_number}
 \begin{split}
 \frac{\partial P(n,t)}{\partial t}
  & =  W(n| n + 1) P(n + 1,t) + W(n| n - 1) P(n - 1,t) \\
  & \qquad - [W(n + 1|n) + W(n - 1|n)] P(n,t) \\
  & = \frac{1}{\tau} [(n + 1) P(n + 1,t) + \bar{n} P(n - 1,t) - (\bar{n} + n) P(n,t)] ,
 \end{split}
\end{equation}
where $W(n - 1| n)$ and $W(n + 1| n)$ are the reconstruction rates
(eqs \eqref{destruction_rate_bonds} and \eqref{construction_rate_bonds}),
{and} $\bar{n}$ is the
average number of the bonds for a fixed bond vector $\bm{R}_{i} -
\bm{R}_{j}$,
\begin{equation}
 \bar{n} \equiv \xi e^{- u(\bm{R}_{i} - \bm{R}_{j}) / k_{B} T}.
\end{equation}
We introduce the difference and averaging operators in the bond number
space:
\begin{align}
 \label{difference_operator}
 \hat{D} f(n) & \equiv f(n + 1/2) - f(n - 1/2) , \\
 \label{averaging_operator}
 \hat{M} f(n) & \equiv \frac{f(n + 1/2) + f(n + 1/2)}{2} ,
\end{align}
where $f(n)$ is a given function of $n$. With the difference operator $\hat{D}$,
eq \eqref{master_equation_bond_number} can be rewritten as follows:
\begin{equation}
 \label{master_equation_bond_number_modified}
 \begin{split}
 \frac{\partial P(n,t)}{\partial t}
  & = - \frac{1}{\tau} [
  [\bar{n}  P(n,t) - (n + 1) P(n + 1,t)] 
  - [\bar{n} P(n - 1,t) - n P(n,t)]
   ] \\
  & = - \hat{D} J(n,t) .
 \end{split}
\end{equation}
Here we have defined the flux in the bond number space, $J(n + 1/2,t)$:
\begin{equation}
 \label{bond_number_flux}
  J(n + 1/2,t)
    \equiv \frac{1}{\tau} [\bar{n} P(n,t) - (n + 1) P(n + 1,t)] .
\end{equation}
Eq \eqref{master_equation_bond_number_modified} has the form of the conservation
equation, and if the flux can be related to the difference of the
probability distribution, the master equation can be expressed as a
Fokker-Planck type equation.
The flux can be rewritten as the following form, by utilizing the
difference and averaging operators:
\begin{equation}
 \label{bond_number_flux_modified}
  \begin{split}
  J(n + 1/2,t)
   & = \frac{1}{\tau} [\bar{n} P(n,t) - (n + 1) P(n + 1,t)] \\
   & = \frac{1}{\tau} [\bar{n} (\hat{M} - \hat{D}/2) P(n + 1/2,t) - (n +
   1) (\hat{M} + \hat{D}/2) P(n + 1/2,t)] \\
%   & = \frac{1}{\tau} \left[ [\bar{n} - (n + 1)] \hat{M} P(n + 1/2,t)
%   - \frac{\bar{n} + (n +  1)}{2} \hat{D} P(n + 1/2,t) \right] \\
   & = - \frac{1}{\tau} \left[ \frac{(n +  1) + \bar{n}}{2}
   \left[ 2 \frac{(n + 1) - \bar{n}}{(n +  1) + \bar{n}} \hat{M} P(n + 1/2,t)
   + \hat{D} P(n + 1/2,t)  \right]
   \right] .
  \end{split}
\end{equation}
By substituting eq \eqref{bond_number_flux_modified} into \eqref{master_equation_bond_number_modified},
the master equation can be rewritten as a Fokker-Planck type equation\cite{vanKampen-book}.
We assume that $n$ is large, and approximate the difference operator
as the differential operator. In addition, {we simply
ignore} the averaging operator. {We expand} the bond number $n$ around its
average value $\bar{n}${, and keep only the leading order terms}. Then, the master equation can be approximated as
\begin{equation}
 \begin{split}
  \label{master_equation_bond_number_approximated}
 \frac{\partial P(n,t)}{\partial t}
  & = \frac{1}{\tau} \hat{D}
 \left[ \frac{(n + 1/2) + \bar{n}}{2}
   \left[ 2 \frac{(n + 1/2) - \bar{n}}{(n +  1/2) + \bar{n}} \hat{M} P(n,t)
   + \hat{D} P(n,t)  \right]
   \right] \\
  & \approx \frac{\bar{n}}{\tau} \frac{\partial}{\partial n}
   \left[ \frac{n - \bar{n}}{\bar{n}} P(n,t)
   + \frac{\partial  P(n,t)}{\partial n}  \right] .
 \end{split}
\end{equation}

Now eq \eqref{master_equation_bond_number_approximated} can be
interpreted as the Fokker-Planck equation for the bond number $n$. The
corresponding Langevin equation for the number of transient bonds between the
$i$-th and $j$-th particles, $n_{ij}$, is
\begin{equation}
 \label{langevin_equation_bond_number}
 \frac{d n_{ij}(t)}{d t} = - \frac{n_{ij} - \bar{n}_{ij}(\bm{R}_{i} - \bm{R}_{j})}{\tau} + \sqrt{\frac{2 \bar{n}_{ij}(\bm{R}_{i} - \bm{R}_{j})}{\tau}} \, w'_{ij}(t) ,
\end{equation}
where $\bar{n}_{ij}(\bm{R}_{i} - \bm{R}_{j}) \equiv \xi \exp(-u(\bm{R}_{i} - \bm{R}_{j}) / k_{B} T)$,
and $w_{ij}'(t)$ is the Gaussian white noise. The first and second
moments of $w_{ij}'(t)$ are given as
\begin{equation}
 \langle w'_{ij}(t) \rangle = 0, \qquad \langle w'_{ij}(t) w'_{kl}(t') \rangle =
  \delta_{ij,kl} \delta(t - t') .
\end{equation}
The Langevin equation for the particle positions (eq
\eqref{langevin_equation_particles}) can be rewritten as follows, in
absence of the external flow ($\bm{\kappa} = 0$):
\begin{equation}
 \label{langevin_equation_particles_approx}
  \frac{d\bm{R}_{i}(t)}{dt} = - \frac{1}{\zeta} \sum_{j}
 \left[ \frac{\partial v(\bm{R}_{i} - \bm{R}_{j})}{\partial \bm{R}_{i}}
  + [n_{ij} - \bar{n}_{ij}(\bm{R}_{i} - \bm{R}_{j})] \frac{\partial u(\bm{R}_{i} - \bm{R}_{j})}{\partial \bm{R}_{j}}
 \right] + \sqrt{\frac{2 k_{B} T}{\zeta}}
  \bm{w}_{i}(t) .
\end{equation}
Also, the equilibrium probability distribution (eq
\eqref{probability_distribution_positions_bonds_equilibrium}) can be approximated by expanding the exponent $n_{ij}$ around the most probable value
$\bar{n}_{ij}(\bm{R}_{i} - \bm{R}_{j})$:
\begin{equation}
 \begin{split}
 \label{probability_distribution_positions_bonds_equilibrium_approx}
  P_{\text{eq}}(\lbrace \bm{R}_{i}
  \rbrace, \lbrace n_{ij} \rbrace) 
  & \approx \frac{1}{\displaystyle \mathcal{Z} \Lambda^{3 M} M!}
  \left[ \prod_{i > j} \frac{1}{\sqrt{2 \pi
  \bar{n}_{ij}(\bm{R}_{i} - \bm{R}_{j})}} \right]
 \\
  & \qquad  \times  \exp\left[
      - \frac{1}{k_{B} T} \sum_{i > j} v(\bm{R}_{i} - \bm{R}_{j})
  - \sum_{i > j} \frac{[n_{ij} - \bar{n}_{ij}(\bm{R}_{i} -
  \bm{R}_{j})]^{2}}{2 \bar{n}_{ij}(\bm{R}_{i} - \bm{R}_{j})} \right] .
 \end{split}
\end{equation}

Eqs \eqref{langevin_equation_particles_approx} 
\eqref{langevin_equation_bond_number}, and
\eqref{probability_distribution_positions_bonds_equilibrium_approx}
have very similar forms to the
Langevin equation for the particle positions, {one for} the numbers of
entanglements, and the probability distribution, in the 
original RaPiD
model. (See eqs (3), (6), and (1) in Ref.~\cite{Kindt-Briels-2007}.)
{In the original RaPiD model, the contribution of the
transient bonds to the equilibrium probability
distribution is modelled by the harmonic free energy for bond numbers:
\begin{equation}
 \label{rapid_bond_potential_energy}
  \mathcal{F}_{\text{bond}}(\lbrace \bm{R}_{i} \rbrace,
  \lbrace n_{ij} \rbrace) =
 \sum_{i > j} \frac{k_{B} T \alpha}{2} [n_{ij} - \bar{n}_{ij}(\bm{R}_{i} - \bm{R}_{j})]^{2}
\end{equation}
where $\alpha$ is assumed to be constant. We can find a similar factor in the exponent
in
eq~\eqref{probability_distribution_positions_bonds_equilibrium_approx}.
If we set $\alpha$ as $\alpha = 1 / \bar{n}_{ij}(\bm{R}_{i} -
\bm{R}_{j})$ (which is not constant but depends on the bond vector) in
eq~\eqref{rapid_bond_potential_energy}, the equilibrium probability distribution
\eqref{probability_distribution_positions_bonds_equilibrium_approx}
becomes the same as one in the RaPiD model.

From the viewpoint of the dynamics, both our model and the
RaPiD model are described by the Langevin equations if the bond number
is large.}
In the RaPiD model, the noise term in the Langevin equation for the
particle positions {generally} depends on the bond vector and the number of
entanglements, whereas the noise term in the Langevin equation for the
numbers of entanglement is independent of the
bond vector. In contrast, the noise term in eq
\eqref{langevin_equation_particles_approx} is independent of the bond
vector and the bond number.
{We can 
interpreted eq~\eqref{langevin_equation_particles_approx} as
the special case of the RaPiD model, where the friction coefficient is
assumed to be constant. On the other hand,
the noise term in eq~\eqref{langevin_equation_bond_number} explicitly
depends on the bond
vector via $\bar{n}_{ij}(\bm{R}_{i} - \bm{R}_{j})$.
In general, the Langevin equation for the bond number should be given as
the following form, with the effective free energy $\mathcal{F}(\lbrace
\bm{R}_{i} \rbrace, \lbrace n_{ij} \rbrace)$ and the mobility $L_{ij}(\lbrace
\bm{R}_{i} \rbrace, \lbrace n_{ij} \rbrace)$:
\begin{equation}
\label{langevin_equation_bond_number_general}
\begin{split}
  \frac{d n_{ij}(t)}{d t} = & - L_{ij}(\lbrace
 \bm{R}_{i} \rbrace, \lbrace n_{ij} \rbrace) \frac{\partial
 \mathcal{F}(\lbrace \bm{R}_{i} \rbrace, \lbrace n_{ij} \rbrace)}{\partial
 n_{ij}} \\
 & + k_{B} T  \frac{\partial  L_{ij}(\lbrace \bm{R}_{i} \rbrace, \lbrace n_{ij} \rbrace)}{\partial n_{ij}} 
 + \sqrt{2 k_{B} T L_{ij}^{(n)}(\lbrace
 \bm{R}_{i} \rbrace, \lbrace n_{ij} \rbrace)} \, w'_{ij}(t) ,
\end{split}
\end{equation}
\begin{equation}
 \label{effective_free_energy_approx}
 \mathcal{F}(\lbrace \bm{R}_{i} \rbrace, \lbrace n_{ij} \rbrace)
  \approx \sum_{i > j} v(\bm{R}_{i} - \bm{R}_{j})
  + \frac{k_{B} T [n_{ij} - \bar{n}_{ij}(\bm{R}_{i} -
  \bm{R}_{j})]^{2}}{2 \bar{n}_{ij}(\bm{R}_{i} - \bm{R}_{j})} .
\end{equation}
Comparing eqs~\eqref{langevin_equation_bond_number_general}
and \eqref{effective_free_energy_approx} with
eq~\eqref{langevin_equation_bond_number}, we find
$L_{ij}(\lbrace
\bm{R}_{i} \rbrace, \lbrace n_{ij} \rbrace) = \bar{n}_{ij}(\bm{R}_{i} -
\bm{R}_{j}) / \tau k_{B} T$.
Again, we can interpret our model as the special case of the RaPiD model
where the friction coefficient for the bond number depends on the
equilibrium bond number.}
Therefore, our model
can be interpreted as a variant of the RaPiD model.

\section{Mean-Field Approximation For Single Particle Dynamics}
\label{mean_field_approximation_for_single_particle_dynamics}

To study the dynamical properties of the transient bond model, we
consider the mean-field approximation. We consider the statics and
dynamics of a single particle in an ideal transient bond {model} without
the external flow.
The static property can be obtained by the equilibrium probability
distribution for a single particle. Under the mean-field approximation,
the information of other particles is smeared out. This situation is
almost the same as the relation between 
the multi-chain slip-spring model and the single-chain slip-spring model
as a mean-field model.

We describe the position of {a} target particle as $\bm{R}$, and the
number of transient bonds attached to the target particle as $n$.
{In} the ideal transient bond {model}, the transient bond is expressed
as the harmonic potentials. One end of the $j$-th transient bond is attached to
the target particle, and we assume that another end is anchored
{in space,} at
$\bm{A}_{j}$.  Then, we can approximately express the equilibrium probability
distribution as follows:
\begin{equation}
 \label{probability_distribution_mean_field}
 P_{\text{eq}}(\bm{R},n,\lbrace \bm{A}_{j} \rbrace)
  = \frac{e^{-\tilde{\xi}}}{V n!} \left(\frac{3}{2 \pi a}\right)^{3 n/2} \tilde{\xi}^{n}
  \exp\left[ - \sum_{j = 1}^{n} \frac{3 }{2 a^{2}} (\bm{R} - \bm{A}_{j})^{2}
      \right],
\end{equation}
where $\tilde{\xi}$ is the effective fugacity, and it does not coincide to
the fugacity $\xi$ of the transient bond model in the main text. From eq
\eqref{probability_distribution_mean_field}, the mean-field free energy
simply becomes the {sum of harmonic bond potentials}. Thus the Langevin equation can be
described as
\begin{equation}
 \label{langevin_equation_mean_field}
 \frac{d \bm{R}(t)}{dt} = - \sum_{j = 1}^{n} \frac{3 k_{B} T}{\zeta a^{2}} (\bm{R} -
  \bm{A}_{j}) + \sqrt{\frac{2 k_{B} T}{\zeta}} \bm{w}(t) ,
\end{equation}
where $\bm{w}(t)$ is the Gaussian white noise which satisfies
\begin{equation}
 \langle \bm{w}(t) \rangle = 0, \qquad
  \langle \bm{w}(t) \bm{w}(t') \rangle = \bm{1} \delta(t - t') .
\end{equation}
The destruction rate of a single transient bond is simply given as
\begin{equation}
 \label{destruction_rate_mean_field}
  W_{-} = \frac{1}{\tau} .
\end{equation}
We assume that the transient bond indices are renumbered after the
destruction.
The construction rate is determined from the detailed-balance condition.
From eqs \eqref{probability_distribution_mean_field} and
\eqref{destruction_rate_mean_field}, we have the construction rate
\begin{equation}
 \label{construction_rate_mean_field}
  W_{+} = \frac{1}{\tau} \tilde{\xi} \left(\frac{3}{2 \pi a}\right)^{3/2}
  \exp\left[ - \frac{3 }{2 a^{2}} (\bm{R} - \bm{A}_{n + 1})^{2}
      \right],
\end{equation}
where $\bm{A}_{n + 1}$ is the position of a new anchoring point.

{Some} dynamical properties of the transient bond model can
be {approximately} analyzed by
the mean-field model shown above. From eqs
\eqref{destruction_rate_mean_field} and
\eqref{construction_rate_mean_field}, the reconstitution process of each transient bond
is independent. Thus we have a single exponential type relaxation
for the stress:
\begin{equation}
 G(t) \approx \rho k_{B} T \tilde{\xi} e^{- t / \tau} .
\end{equation}
The particle is effectively trapped at the average anchoring position
unless $n = 0$. Thus, for the constrained state ($n \ge 1$), we may rewrite eq \eqref{langevin_equation_mean_field} as
\begin{equation}
 \label{langevin_equation_mean_field_modified}
 \frac{d \bm{R}(t)}{dt} = - \frac{3 n k_{B} T}{\zeta a^{2}} (\bm{R} -
  \bar{\bm{A}}) + \sqrt{\frac{2 k_{B} T}{\zeta}} \bm{w}(t) ,
\end{equation}
with the average anchoring position
\begin{equation}
 \bar{\bm{A}} \equiv \frac{1}{n} \sum_{j = 1}^{n} \bm{A}_{j} .
\end{equation}
At the short time scale, the particle can freely diffuse but the
particle is trapped around the anchoring point, thus at the moderate
time scale the motion is strongly constrained.
At the long time scale, the particle motion is almost the same as the motion
of the average anchoring position ($\bm{R} \approx \bar{\bm{A}}$).
Therefore, the mean-square displacement shows the transition from the
normal diffusion to subdiffusion, and then to the normal diffusion.
The average anchoring position will
move roughly by $a / n$ by one reconstruction. If we replace the number
of transient bonds by its equilibrium average, we have $a / n \approx a
/ \tilde{\xi}$. Then, the diffusion
coefficient at the constrained state is roughly estimated as
\begin{equation}
 D_{\text{constrained}} \propto \frac{a^{2}}{\tilde{\xi}^{2} \tau} .
\end{equation}
If the number of bonds is zero, $n = 0$, the particle is at the free state
and thus it can freely diffuse. The diffusion coefficient at this free
state is thus simply estimated as
\begin{equation}
 D_{\text{free}} = \frac{k_{B} T}{\zeta} .
\end{equation}
The diffusion coefficient is given as the average of the diffusion
coefficients at the {free} and {constrained} states:
\begin{equation}
 \label{diffusion_coefficient_mean_field}
 D = e^{-\tilde{\xi}} D_{\text{free}}  + (1 - e^{-\tilde{\xi}})  D_{\text{constrained}}.
\end{equation}
There are two extreme cases where the diffusion coefficient becomes simple.
If the effective fugacity $\tilde{\xi}$ is sufficiently large, the first
term in {the right hand side of} eq \eqref{diffusion_coefficient_mean_field} becomes negligible
and we have $D \approx D_{\text{constrained}}$. 
If $\tau$ is sufficiently large, the second term in {the right hand side of} eq \eqref{diffusion_coefficient_mean_field} becomes negligible and
we have $D \approx e^{-\tilde{\xi}} D_{\text{free}}$, and $D$ becomes
independent of $\tau$.
Although these estimates {can} not fully explain the simulation results for
the transient bond model, we can understand some aspects of the dynamical
properties of the transient bond model.

%------------------------------------------------------------------------------
\bibliographystyle{apsrev4-1}
\bibliography{transient_bond_model}

\begin{thebibliography}{44}%
\makeatletter
\providecommand \@ifxundefined [1]{%
 \@ifx{#1\undefined}
}%
\providecommand \@ifnum [1]{%
 \ifnum #1\expandafter \@firstoftwo
 \else \expandafter \@secondoftwo
 \fi
}%
\providecommand \@ifx [1]{%
 \ifx #1\expandafter \@firstoftwo
 \else \expandafter \@secondoftwo
 \fi
}%
\providecommand \natexlab [1]{#1}%
\providecommand \enquote  [1]{``#1''}%
\providecommand \bibnamefont  [1]{#1}%
\providecommand \bibfnamefont [1]{#1}%
\providecommand \citenamefont [1]{#1}%
\providecommand \href@noop [0]{\@secondoftwo}%
\providecommand \href [0]{\begingroup \@sanitize@url \@href}%
\providecommand \@href[1]{\@@startlink{#1}\@@href}%
\providecommand \@@href[1]{\endgroup#1\@@endlink}%
\providecommand \@sanitize@url [0]{\catcode `\\12\catcode `\$12\catcode
  `\&12\catcode `\#12\catcode `\^12\catcode `\_12\catcode `\%12\relax}%
\providecommand \@@startlink[1]{}%
\providecommand \@@endlink[0]{}%
\providecommand \url  [0]{\begingroup\@sanitize@url \@url }%
\providecommand \@url [1]{\endgroup\@href {#1}{\urlprefix }}%
\providecommand \urlprefix  [0]{URL }%
\providecommand \Eprint [0]{\href }%
\providecommand \doibase [0]{http://dx.doi.org/}%
\providecommand \selectlanguage [0]{\@gobble}%
\providecommand \bibinfo  [0]{\@secondoftwo}%
\providecommand \bibfield  [0]{\@secondoftwo}%
\providecommand \translation [1]{[#1]}%
\providecommand \BibitemOpen [0]{}%
\providecommand \bibitemStop [0]{}%
\providecommand \bibitemNoStop [0]{.\EOS\space}%
\providecommand \EOS [0]{\spacefactor3000\relax}%
\providecommand \BibitemShut  [1]{\csname bibitem#1\endcsname}%
\let\auto@bib@innerbib\@empty
%</preamble>
\bibitem [{\citenamefont {Doi}\ and\ \citenamefont
  {Edwards}(1986)}]{Doi-Edwards-book}%
  \BibitemOpen
  \bibfield  {author} {\bibinfo {author} {\bibfnamefont {M.}~\bibnamefont
  {Doi}}\ and\ \bibinfo {author} {\bibfnamefont {S.~F.}\ \bibnamefont
  {Edwards}},\ }\href@noop {} {\emph {\bibinfo {title} {The Theory of Polymer
  Dynamics}}}\ (\bibinfo  {publisher} {Oxford University Press},\ \bibinfo
  {address} {Oxford},\ \bibinfo {year} {1986})\BibitemShut {NoStop}%
\bibitem [{\citenamefont {Kremer}\ and\ \citenamefont
  {Grest}(1990)}]{Kremer-Grest-1990}%
  \BibitemOpen
  \bibfield  {author} {\bibinfo {author} {\bibfnamefont {K.}~\bibnamefont
  {Kremer}}\ and\ \bibinfo {author} {\bibfnamefont {G.~S.}\ \bibnamefont
  {Grest}},\ }\href@noop {} {\bibfield  {journal} {\bibinfo  {journal} {J.
  Chem. Phys.}\ }\textbf {\bibinfo {volume} {92}},\ \bibinfo {pages} {5057}
  (\bibinfo {year} {1990})}\BibitemShut {NoStop}%
\bibitem [{\citenamefont {Hua}\ and\ \citenamefont
  {Schieber}(1998)}]{Hua-Schieber-1998}%
  \BibitemOpen
  \bibfield  {author} {\bibinfo {author} {\bibfnamefont {C.~C.}\ \bibnamefont
  {Hua}}\ and\ \bibinfo {author} {\bibfnamefont {J.~D.}\ \bibnamefont
  {Schieber}},\ }\href@noop {} {\bibfield  {journal} {\bibinfo  {journal} {J.
  Chem. Phys.}\ }\textbf {\bibinfo {volume} {109}},\ \bibinfo {pages} {10018}
  (\bibinfo {year} {1998})}\BibitemShut {NoStop}%
\bibitem [{\citenamefont {Masubuchi}\ \emph {et~al.}(2001)\citenamefont
  {Masubuchi}, \citenamefont {Takimoto}, \citenamefont {Koyama}, \citenamefont
  {Ianniruberto}, \citenamefont {Greco},\ and\ \citenamefont
  {Marrucci}}]{Masubuchi-Takimoto-Koyama-Ianniruberto-Greco-Marrucci-2001}%
  \BibitemOpen
  \bibfield  {author} {\bibinfo {author} {\bibfnamefont {Y.}~\bibnamefont
  {Masubuchi}}, \bibinfo {author} {\bibfnamefont {J.}~\bibnamefont {Takimoto}},
  \bibinfo {author} {\bibfnamefont {K.}~\bibnamefont {Koyama}}, \bibinfo
  {author} {\bibfnamefont {G.}~\bibnamefont {Ianniruberto}}, \bibinfo {author}
  {\bibfnamefont {F.}~\bibnamefont {Greco}}, \ and\ \bibinfo {author}
  {\bibfnamefont {G.}~\bibnamefont {Marrucci}},\ }\href@noop {} {\bibfield
  {journal} {\bibinfo  {journal} {J. Chem. Phys.}\ }\textbf {\bibinfo {volume}
  {115}},\ \bibinfo {pages} {4387} (\bibinfo {year} {2001})}\BibitemShut
  {NoStop}%
\bibitem [{\citenamefont {Schieber}(2003)}]{Schieber-2003}%
  \BibitemOpen
  \bibfield  {author} {\bibinfo {author} {\bibfnamefont {J.~D.}\ \bibnamefont
  {Schieber}},\ }\href@noop {} {\bibfield  {journal} {\bibinfo  {journal} {J.
  Chem. Phys.}\ }\textbf {\bibinfo {volume} {118}},\ \bibinfo {pages} {5162}
  (\bibinfo {year} {2003})}\BibitemShut {NoStop}%
\bibitem [{\citenamefont {Doi}\ and\ \citenamefont
  {Takimoto}(2003)}]{Doi-Takimoto-2003}%
  \BibitemOpen
  \bibfield  {author} {\bibinfo {author} {\bibfnamefont {M.}~\bibnamefont
  {Doi}}\ and\ \bibinfo {author} {\bibfnamefont {J.}~\bibnamefont {Takimoto}},\
  }\href@noop {} {\bibfield  {journal} {\bibinfo  {journal} {Phil. Trans. R.
  Soc. Lond. A}\ }\textbf {\bibinfo {volume} {361}},\ \bibinfo {pages} {641}
  (\bibinfo {year} {2003})}\BibitemShut {NoStop}%
\bibitem [{\citenamefont {Nair}\ and\ \citenamefont
  {Schieber}(2006)}]{Nair-Schieber-2006}%
  \BibitemOpen
  \bibfield  {author} {\bibinfo {author} {\bibfnamefont {D.~M.}\ \bibnamefont
  {Nair}}\ and\ \bibinfo {author} {\bibfnamefont {J.~D.}\ \bibnamefont
  {Schieber}},\ }\href@noop {} {\bibfield  {journal} {\bibinfo  {journal}
  {Macromolecules}\ }\textbf {\bibinfo {volume} {39}},\ \bibinfo {pages} {3386}
  (\bibinfo {year} {2006})}\BibitemShut {NoStop}%
\bibitem [{\citenamefont {Khaliullin}\ and\ \citenamefont
  {Schieber}(2009)}]{Khaliullin-Schieber-2009}%
  \BibitemOpen
  \bibfield  {author} {\bibinfo {author} {\bibfnamefont {R.~N.}\ \bibnamefont
  {Khaliullin}}\ and\ \bibinfo {author} {\bibfnamefont {J.~D.}\ \bibnamefont
  {Schieber}},\ }\href@noop {} {\bibfield  {journal} {\bibinfo  {journal}
  {Macromolecules}\ }\textbf {\bibinfo {volume} {42}},\ \bibinfo {pages} {7504}
  (\bibinfo {year} {2009})}\BibitemShut {NoStop}%
\bibitem [{\citenamefont {Likhtman}(2005)}]{Likhtman-2005}%
  \BibitemOpen
  \bibfield  {author} {\bibinfo {author} {\bibfnamefont {A.~E.}\ \bibnamefont
  {Likhtman}},\ }\href@noop {} {\bibfield  {journal} {\bibinfo  {journal}
  {Macromolecules}\ }\textbf {\bibinfo {volume} {38}},\ \bibinfo {pages} {6128}
  (\bibinfo {year} {2005})}\BibitemShut {NoStop}%
\bibitem [{\citenamefont {Uneyama}(2011)}]{Uneyama-2011}%
  \BibitemOpen
  \bibfield  {author} {\bibinfo {author} {\bibfnamefont {T.}~\bibnamefont
  {Uneyama}},\ }\href@noop {} {\bibfield  {journal} {\bibinfo  {journal} {Nihon
  Reoroji Gakkaishi (J. Soc. Rheol. Jpn.)}\ }\textbf {\bibinfo {volume} {39}},\
  \bibinfo {pages} {135} (\bibinfo {year} {2011})}\BibitemShut {NoStop}%
\bibitem [{\citenamefont {Chappa}\ \emph {et~al.}(2012)\citenamefont {Chappa},
  \citenamefont {Morse}, \citenamefont {Zippelius},\ and\ \citenamefont
  {M\"{u}ller}}]{Chappa-Morse-Zippelius-Muller-2012}%
  \BibitemOpen
  \bibfield  {author} {\bibinfo {author} {\bibfnamefont {V.~C.}\ \bibnamefont
  {Chappa}}, \bibinfo {author} {\bibfnamefont {D.~C.}\ \bibnamefont {Morse}},
  \bibinfo {author} {\bibfnamefont {A.}~\bibnamefont {Zippelius}}, \ and\
  \bibinfo {author} {\bibfnamefont {M.}~\bibnamefont {M\"{u}ller}},\
  }\href@noop {} {\bibfield  {journal} {\bibinfo  {journal} {Phys. Rev. Lett.}\
  }\textbf {\bibinfo {volume} {109}},\ \bibinfo {pages} {148302} (\bibinfo
  {year} {2012})}\BibitemShut {NoStop}%
\bibitem [{\citenamefont {Uneyama}\ and\ \citenamefont
  {Masubuchi}(2012)}]{Uneyama-Masubuchi-2012}%
  \BibitemOpen
  \bibfield  {author} {\bibinfo {author} {\bibfnamefont {T.}~\bibnamefont
  {Uneyama}}\ and\ \bibinfo {author} {\bibfnamefont {Y.}~\bibnamefont
  {Masubuchi}},\ }\href@noop {} {\bibfield  {journal} {\bibinfo  {journal} {J.
  Chem. Phys.}\ }\textbf {\bibinfo {volume} {137}},\ \bibinfo {pages} {154902}
  (\bibinfo {year} {2012})}\BibitemShut {NoStop}%
\bibitem [{\citenamefont {Kindt}\ and\ \citenamefont
  {Briels}(2007)}]{Kindt-Briels-2007}%
  \BibitemOpen
  \bibfield  {author} {\bibinfo {author} {\bibfnamefont {P.}~\bibnamefont
  {Kindt}}\ and\ \bibinfo {author} {\bibfnamefont {W.~J.}\ \bibnamefont
  {Briels}},\ }\href@noop {} {\bibfield  {journal} {\bibinfo  {journal} {J.
  Chem. Phys.}\ }\textbf {\bibinfo {volume} {127}},\ \bibinfo {pages} {134901}
  (\bibinfo {year} {2007})}\BibitemShut {NoStop}%
\bibitem [{\citenamefont {{Santos de Oliveira}}\ \emph
  {et~al.}(2014)\citenamefont {{Santos de Oliveira}}, \citenamefont
  {Fitzgerald}, , \citenamefont {den Otter},\ and\ \citenamefont
  {Briels}}]{SantosdeOliveira-Fitzgerald-denOtter-Briels-2014}%
  \BibitemOpen
  \bibfield  {author} {\bibinfo {author} {\bibfnamefont {I.~S.}\ \bibnamefont
  {{Santos de Oliveira}}}, \bibinfo {author} {\bibfnamefont {B.~W.}\
  \bibnamefont {Fitzgerald}}, , \bibinfo {author} {\bibfnamefont {W.~K.}\
  \bibnamefont {den Otter}}, \ and\ \bibinfo {author} {\bibfnamefont {W.~J.}\
  \bibnamefont {Briels}},\ }\href@noop {} {\bibfield  {journal} {\bibinfo
  {journal} {J. Chem. Phys.}\ }\textbf {\bibinfo {volume} {140}},\ \bibinfo
  {pages} {104903} (\bibinfo {year} {2014})}\BibitemShut {NoStop}%
\bibitem [{\citenamefont {Sprakel}\ \emph {et~al.}(2009)\citenamefont
  {Sprakel}, \citenamefont {Spruijt}, \citenamefont {van~der Gucht},
  \citenamefont {Padding},\ and\ \citenamefont
  {Briels}}]{Sprakel-Spruijt-vanderGucht-Padding-Briels-2009}%
  \BibitemOpen
  \bibfield  {author} {\bibinfo {author} {\bibfnamefont {J.}~\bibnamefont
  {Sprakel}}, \bibinfo {author} {\bibfnamefont {E.}~\bibnamefont {Spruijt}},
  \bibinfo {author} {\bibfnamefont {J.}~\bibnamefont {van~der Gucht}}, \bibinfo
  {author} {\bibfnamefont {J.~T.}\ \bibnamefont {Padding}}, \ and\ \bibinfo
  {author} {\bibfnamefont {W.~J.}\ \bibnamefont {Briels}},\ }\href@noop {}
  {\bibfield  {journal} {\bibinfo  {journal} {Soft Matter}\ }\textbf {\bibinfo
  {volume} {5}},\ \bibinfo {pages} {4748} (\bibinfo {year} {2009})}\BibitemShut
  {NoStop}%
\bibitem [{\citenamefont {Sprakel}\ \emph {et~al.}(2011)\citenamefont
  {Sprakel}, \citenamefont {Padding},\ and\ \citenamefont
  {Briels}}]{Sprakel-Padding-Briels-2011}%
  \BibitemOpen
  \bibfield  {author} {\bibinfo {author} {\bibfnamefont {J.}~\bibnamefont
  {Sprakel}}, \bibinfo {author} {\bibfnamefont {J.~T.}\ \bibnamefont
  {Padding}}, \ and\ \bibinfo {author} {\bibfnamefont {W.~J.}\ \bibnamefont
  {Briels}},\ }\href@noop {} {\bibfield  {journal} {\bibinfo  {journal}
  {Europhys. Lett.}\ }\textbf {\bibinfo {volume} {93}},\ \bibinfo {pages}
  {58003} (\bibinfo {year} {2011})}\BibitemShut {NoStop}%
\bibitem [{\citenamefont {Liu}\ \emph {et~al.}(2013)\citenamefont {Liu},
  \citenamefont {Padding}, \citenamefont {den Otter},\ and\ \citenamefont
  {Briels}}]{Liu-Padding-denOtter-Briels-2013}%
  \BibitemOpen
  \bibfield  {author} {\bibinfo {author} {\bibfnamefont {L.}~\bibnamefont
  {Liu}}, \bibinfo {author} {\bibfnamefont {J.~T.}\ \bibnamefont {Padding}},
  \bibinfo {author} {\bibfnamefont {W.~K.}\ \bibnamefont {den Otter}}, \ and\
  \bibinfo {author} {\bibfnamefont {W.~J.}\ \bibnamefont {Briels}},\
  }\href@noop {} {\bibfield  {journal} {\bibinfo  {journal} {J. Chem. Phys.}\
  }\textbf {\bibinfo {volume} {138}},\ \bibinfo {pages} {244912} (\bibinfo
  {year} {2013})}\BibitemShut {NoStop}%
\bibitem [{\citenamefont {Liu}\ \emph {et~al.}(2014)\citenamefont {Liu},
  \citenamefont {den Otter},\ and\ \citenamefont
  {Briels}}]{Liu-denOtter-Briels-2014}%
  \BibitemOpen
  \bibfield  {author} {\bibinfo {author} {\bibfnamefont {L.}~\bibnamefont
  {Liu}}, \bibinfo {author} {\bibfnamefont {W.~K.}\ \bibnamefont {den Otter}},
  \ and\ \bibinfo {author} {\bibfnamefont {W.~J.}\ \bibnamefont {Briels}},\
  }\href@noop {} {\bibfield  {journal} {\bibinfo  {journal} {Soft Matter}\
  }\textbf {\bibinfo {volume} {10}},\ \bibinfo {pages} {7874} (\bibinfo {year}
  {2014})}\BibitemShut {NoStop}%
\bibitem [{\citenamefont {Fitzgerald}\ \emph {et~al.}(2014)\citenamefont
  {Fitzgerald}, \citenamefont {Lentzakis}, \citenamefont {Sakellariou},
  \citenamefont {Vlassopoulos},\ and\ \citenamefont
  {Briels}}]{Fitzgerald-Lentzakis-Sakellariou-Vlassopoulos-Briels-2014}%
  \BibitemOpen
  \bibfield  {author} {\bibinfo {author} {\bibfnamefont {B.~W.}\ \bibnamefont
  {Fitzgerald}}, \bibinfo {author} {\bibfnamefont {H.}~\bibnamefont
  {Lentzakis}}, \bibinfo {author} {\bibfnamefont {G.}~\bibnamefont
  {Sakellariou}}, \bibinfo {author} {\bibfnamefont {D.}~\bibnamefont
  {Vlassopoulos}}, \ and\ \bibinfo {author} {\bibfnamefont {W.~J.}\
  \bibnamefont {Briels}},\ }\href@noop {} {\bibfield  {journal} {\bibinfo
  {journal} {J. Chem. Phys.}\ }\textbf {\bibinfo {volume} {141}},\ \bibinfo
  {pages} {114907} (\bibinfo {year} {2014})}\BibitemShut {NoStop}%
\bibitem [{\citenamefont {Fitzgerald}\ and\ \citenamefont
  {Briels}(2018)}]{Fitzgerald-Briels-2018}%
  \BibitemOpen
  \bibfield  {author} {\bibinfo {author} {\bibfnamefont {B.~W.}\ \bibnamefont
  {Fitzgerald}}\ and\ \bibinfo {author} {\bibfnamefont {W.~J.}\ \bibnamefont
  {Briels}},\ }\href@noop {} {\bibfield  {journal} {\bibinfo  {journal}
  {Macromol. Theory Simul.}\ }\textbf {\bibinfo {volume} {27}},\ \bibinfo
  {pages} {1700069} (\bibinfo {year} {2018})}\BibitemShut {NoStop}%
\bibitem [{\citenamefont {Uneyama}\ and\ \citenamefont
  {Horio}(2011)}]{Uneyama-Horio-2011}%
  \BibitemOpen
  \bibfield  {author} {\bibinfo {author} {\bibfnamefont {T.}~\bibnamefont
  {Uneyama}}\ and\ \bibinfo {author} {\bibfnamefont {K.}~\bibnamefont
  {Horio}},\ }\href@noop {} {\bibfield  {journal} {\bibinfo  {journal} {J.
  Polym. Sci. B: Polym. Phys.}\ }\textbf {\bibinfo {volume} {49}},\ \bibinfo
  {pages} {966} (\bibinfo {year} {2011})}\BibitemShut {NoStop}%
\bibitem [{\citenamefont {van Kampen}(2007)}]{vanKampen-book}%
  \BibitemOpen
  \bibfield  {author} {\bibinfo {author} {\bibfnamefont {N.~G.}\ \bibnamefont
  {van Kampen}},\ }\href@noop {} {\emph {\bibinfo {title} {Stochastic Processes
  in Physics and Chemistry}}},\ \bibinfo {edition} {3rd}\ ed.\ (\bibinfo
  {publisher} {Elsevier},\ \bibinfo {address} {Amsterdam},\ \bibinfo {year}
  {2007})\BibitemShut {NoStop}%
\bibitem [{\citenamefont {Ramirez}\ \emph {et~al.}(2007)\citenamefont
  {Ramirez}, \citenamefont {Sukumaran},\ and\ \citenamefont
  {Likhtman}}]{Ramirez-Sukumaran-Likhtman-2007}%
  \BibitemOpen
  \bibfield  {author} {\bibinfo {author} {\bibfnamefont {J.}~\bibnamefont
  {Ramirez}}, \bibinfo {author} {\bibfnamefont {S.~K.}\ \bibnamefont
  {Sukumaran}}, \ and\ \bibinfo {author} {\bibfnamefont {A.~E.}\ \bibnamefont
  {Likhtman}},\ }\href@noop {} {\bibfield  {journal} {\bibinfo  {journal} {J.
  Chem. Phys.}\ }\textbf {\bibinfo {volume} {126}},\ \bibinfo {pages} {244904}
  (\bibinfo {year} {2007})}\BibitemShut {NoStop}%
\bibitem [{\citenamefont {Inoue}\ and\ \citenamefont
  {Osaki}(1996)}]{Inoue-Osaki-1996}%
  \BibitemOpen
  \bibfield  {author} {\bibinfo {author} {\bibfnamefont {T.}~\bibnamefont
  {Inoue}}\ and\ \bibinfo {author} {\bibfnamefont {K.}~\bibnamefont {Osaki}},\
  }\href@noop {} {\bibfield  {journal} {\bibinfo  {journal} {Macromolecules}\
  }\textbf {\bibinfo {volume} {29}},\ \bibinfo {pages} {1595} (\bibinfo {year}
  {1996})}\BibitemShut {NoStop}%
\bibitem [{\citenamefont {Kr\"{o}ger}\ \emph {et~al.}(1997)\citenamefont
  {Kr\"{o}ger}, \citenamefont {Luap},\ and\ \citenamefont
  {Muller}}]{Kroger-Luap-Muller-1997}%
  \BibitemOpen
  \bibfield  {author} {\bibinfo {author} {\bibfnamefont {M.}~\bibnamefont
  {Kr\"{o}ger}}, \bibinfo {author} {\bibfnamefont {C.}~\bibnamefont {Luap}}, \
  and\ \bibinfo {author} {\bibfnamefont {R.}~\bibnamefont {Muller}},\
  }\href@noop {} {\bibfield  {journal} {\bibinfo  {journal} {Macromolecules}\
  }\textbf {\bibinfo {volume} {30}},\ \bibinfo {pages} {526} (\bibinfo {year}
  {1997})}\BibitemShut {NoStop}%
\bibitem [{\citenamefont {Evans}\ and\ \citenamefont
  {Morris}(2008)}]{Evans-Morris-book}%
  \BibitemOpen
  \bibfield  {author} {\bibinfo {author} {\bibfnamefont {D.~J.}\ \bibnamefont
  {Evans}}\ and\ \bibinfo {author} {\bibfnamefont {G.~P.}\ \bibnamefont
  {Morris}},\ }\href@noop {} {\emph {\bibinfo {title} {Statistical Mechanics of
  Nonequilibrium Liquids}}},\ \bibinfo {edition} {2nd}\ ed.\ (\bibinfo
  {publisher} {Cambridge University Press},\ \bibinfo {address} {Cambridge},\
  \bibinfo {year} {2008})\BibitemShut {NoStop}%
\bibitem [{\citenamefont {Honeycutt}(1992)}]{Honeycutt-1992}%
  \BibitemOpen
  \bibfield  {author} {\bibinfo {author} {\bibfnamefont {R.~L.}\ \bibnamefont
  {Honeycutt}},\ }\href@noop {} {\bibfield  {journal} {\bibinfo  {journal}
  {Phys. Rev. A}\ }\textbf {\bibinfo {volume} {45}},\ \bibinfo {pages} {600}
  (\bibinfo {year} {1992})}\BibitemShut {NoStop}%
\bibitem [{\citenamefont {Likhtman}(2012)}]{Likhtman-chapter}%
  \BibitemOpen
  \bibfield  {author} {\bibinfo {author} {\bibfnamefont {A.~E.}\ \bibnamefont
  {Likhtman}},\ }in\ \href@noop {} {\emph {\bibinfo {booktitle} {Polymer
  Science: A Comprehensive Reference}}},\ \bibinfo {editor} {edited by\
  \bibinfo {editor} {\bibfnamefont {K.}~\bibnamefont {Matyjaszewski}}\ and\
  \bibinfo {editor} {\bibfnamefont {M.}~\bibnamefont {M\"{o}eller}}}\ (\bibinfo
   {publisher} {Elsevier},\ \bibinfo {address} {Amsterdam},\ \bibinfo {year}
  {2012})\ pp.\ \bibinfo {pages} {133--179}\BibitemShut {NoStop}%
\bibitem [{\citenamefont {Matsumoto}\ and\ \citenamefont
  {Nishimura}(1998)}]{Matsumoto-Nishimura-1998}%
  \BibitemOpen
  \bibfield  {author} {\bibinfo {author} {\bibfnamefont {M.}~\bibnamefont
  {Matsumoto}}\ and\ \bibinfo {author} {\bibfnamefont {T.}~\bibnamefont
  {Nishimura}},\ }\href@noop {} {\bibfield  {journal} {\bibinfo  {journal} {ACM
  Trans. Model. Comp. Simul.}\ }\textbf {\bibinfo {volume} {8}},\ \bibinfo
  {pages} {3} (\bibinfo {year} {1998})},\ \bibinfo {note}
  {http://www.math.sci.hiroshima-u.ac.jp/\~{}m-mat/MT/emt.html}\BibitemShut
  {NoStop}%
\bibitem [{\citenamefont {Likhtman}\ \emph {et~al.}(2007)\citenamefont
  {Likhtman}, \citenamefont {Sukumaran},\ and\ \citenamefont
  {Ramirez}}]{Likhtman-Sukumaran-Ramirez-2007}%
  \BibitemOpen
  \bibfield  {author} {\bibinfo {author} {\bibfnamefont {A.~E.}\ \bibnamefont
  {Likhtman}}, \bibinfo {author} {\bibfnamefont {S.~K.}\ \bibnamefont
  {Sukumaran}}, \ and\ \bibinfo {author} {\bibfnamefont {J.}~\bibnamefont
  {Ramirez}},\ }\href@noop {} {\bibfield  {journal} {\bibinfo  {journal}
  {Macromolecules}\ }\textbf {\bibinfo {volume} {40}},\ \bibinfo {pages} {6748}
  (\bibinfo {year} {2007})}\BibitemShut {NoStop}%
\bibitem [{\citenamefont {Lodge}(1999)}]{Lodge-1999}%
  \BibitemOpen
  \bibfield  {author} {\bibinfo {author} {\bibfnamefont {T.~P.}\ \bibnamefont
  {Lodge}},\ }\href@noop {} {\bibfield  {journal} {\bibinfo  {journal} {Phys.
  Rev. Lett.}\ }\textbf {\bibinfo {volume} {83}},\ \bibinfo {pages} {3218}
  (\bibinfo {year} {1999})}\BibitemShut {NoStop}%
\bibitem [{\citenamefont {Kr\"{o}ger}(2004)}]{Kroger-2004}%
  \BibitemOpen
  \bibfield  {author} {\bibinfo {author} {\bibfnamefont {M.}~\bibnamefont
  {Kr\"{o}ger}},\ }\href@noop {} {\bibfield  {journal} {\bibinfo  {journal}
  {Phys. Rep.}\ }\textbf {\bibinfo {volume} {390}},\ \bibinfo {pages} {453}
  (\bibinfo {year} {2004})}\BibitemShut {NoStop}%
\bibitem [{\citenamefont {Yamamoto}\ and\ \citenamefont
  {Onuki}(1998)}]{Yamamoto-Onuki-1998}%
  \BibitemOpen
  \bibfield  {author} {\bibinfo {author} {\bibfnamefont {R.}~\bibnamefont
  {Yamamoto}}\ and\ \bibinfo {author} {\bibfnamefont {A.}~\bibnamefont
  {Onuki}},\ }\href@noop {} {\bibfield  {journal} {\bibinfo  {journal} {Phys.
  Rev. E}\ }\textbf {\bibinfo {volume} {58}},\ \bibinfo {pages} {3515}
  (\bibinfo {year} {1998})}\BibitemShut {NoStop}%
\bibitem [{\citenamefont {Sillescu}(1999)}]{Sillescu-1999}%
  \BibitemOpen
  \bibfield  {author} {\bibinfo {author} {\bibfnamefont {H.}~\bibnamefont
  {Sillescu}},\ }\href@noop {} {\bibfield  {journal} {\bibinfo  {journal} {J.
  Non-Cryst. Solids}\ }\textbf {\bibinfo {volume} {243}},\ \bibinfo {pages}
  {81} (\bibinfo {year} {1999})}\BibitemShut {NoStop}%
\bibitem [{\citenamefont {Uneyama}\ \emph {et~al.}(2015)\citenamefont
  {Uneyama}, \citenamefont {Miyaguchi},\ and\ \citenamefont
  {Akimoto}}]{Uneyama-Miyaguchi-Akimoto-2015}%
  \BibitemOpen
  \bibfield  {author} {\bibinfo {author} {\bibfnamefont {T.}~\bibnamefont
  {Uneyama}}, \bibinfo {author} {\bibfnamefont {T.}~\bibnamefont {Miyaguchi}},
  \ and\ \bibinfo {author} {\bibfnamefont {T.}~\bibnamefont {Akimoto}},\
  }\href@noop {} {\bibfield  {journal} {\bibinfo  {journal} {Phys. Rev. E}\
  }\textbf {\bibinfo {volume} {92}},\ \bibinfo {pages} {032140} (\bibinfo
  {year} {2015})}\BibitemShut {NoStop}%
\bibitem [{\citenamefont {Liu}\ \emph {et~al.}()\citenamefont {Liu},
  \citenamefont {Halasa}, \citenamefont {Keunings},\ and\ \citenamefont
  {Bailly}}]{Liu-Halasa-Keunings-Bailly-2006}%
  \BibitemOpen
  \bibfield  {author} {\bibinfo {author} {\bibfnamefont {C.-Y.}\ \bibnamefont
  {Liu}}, \bibinfo {author} {\bibfnamefont {A.~F.}\ \bibnamefont {Halasa}},
  \bibinfo {author} {\bibfnamefont {R.}~\bibnamefont {Keunings}}, \ and\
  \bibinfo {author} {\bibfnamefont {C.}~\bibnamefont {Bailly}},\ }\href@noop {}
  {\bibfield  {journal} {\bibinfo  {journal} {Macromolecules}\ }\textbf
  {\bibinfo {volume} {39}},\ \bibinfo {pages} {7415}}\BibitemShut {NoStop}%
\bibitem [{\citenamefont {Matsumiya}\ \emph {et~al.}(2013)\citenamefont
  {Matsumiya}, \citenamefont {Kumazawa}, \citenamefont {Nagao}, \citenamefont
  {Urakawa},\ and\ \citenamefont
  {Watanabe}}]{Matsumiya-Kumazawa-Nagao-Urakawa-Watanabe-2013}%
  \BibitemOpen
  \bibfield  {author} {\bibinfo {author} {\bibfnamefont {Y.}~\bibnamefont
  {Matsumiya}}, \bibinfo {author} {\bibfnamefont {K.}~\bibnamefont {Kumazawa}},
  \bibinfo {author} {\bibfnamefont {M.}~\bibnamefont {Nagao}}, \bibinfo
  {author} {\bibfnamefont {O.}~\bibnamefont {Urakawa}}, \ and\ \bibinfo
  {author} {\bibfnamefont {H.}~\bibnamefont {Watanabe}},\ }\href@noop {}
  {\bibfield  {journal} {\bibinfo  {journal} {Macromolecules}\ }\textbf
  {\bibinfo {volume} {46}},\ \bibinfo {pages} {6067} (\bibinfo {year}
  {2013})}\BibitemShut {NoStop}%
\bibitem [{\citenamefont {Masubuchi}\ \emph {et~al.}(2017)\citenamefont
  {Masubuchi}, \citenamefont {Amamoto}, \citenamefont {Pandey},\ and\
  \citenamefont {Liu}}]{Masubuchi-Amamoto-Pandey-Liu-2017}%
  \BibitemOpen
  \bibfield  {author} {\bibinfo {author} {\bibfnamefont {Y.}~\bibnamefont
  {Masubuchi}}, \bibinfo {author} {\bibfnamefont {Y.}~\bibnamefont {Amamoto}},
  \bibinfo {author} {\bibfnamefont {A.}~\bibnamefont {Pandey}}, \ and\ \bibinfo
  {author} {\bibfnamefont {C.-Y.}\ \bibnamefont {Liu}},\ }\href@noop {}
  {\bibfield  {journal} {\bibinfo  {journal} {Soft Matter}\ }\textbf {\bibinfo
  {volume} {13}},\ \bibinfo {pages} {6585} (\bibinfo {year}
  {2017})}\BibitemShut {NoStop}%
\bibitem [{\citenamefont {Uneyama}\ \emph {et~al.}(2012)\citenamefont
  {Uneyama}, \citenamefont {Suzuki},\ and\ \citenamefont
  {Watanabe}}]{Uneyama-Suzuki-Watanabe-2012}%
  \BibitemOpen
  \bibfield  {author} {\bibinfo {author} {\bibfnamefont {T.}~\bibnamefont
  {Uneyama}}, \bibinfo {author} {\bibfnamefont {S.}~\bibnamefont {Suzuki}}, \
  and\ \bibinfo {author} {\bibfnamefont {H.}~\bibnamefont {Watanabe}},\
  }\href@noop {} {\bibfield  {journal} {\bibinfo  {journal} {Phys. Rev. E}\
  }\textbf {\bibinfo {volume} {86}},\ \bibinfo {pages} {031802} (\bibinfo
  {year} {2012})}\BibitemShut {NoStop}%
\bibitem [{\citenamefont {Masubuchi}\ and\ \citenamefont
  {Uneyama}(2018)}]{Masubuchi-Uneyama-2018}%
  \BibitemOpen
  \bibfield  {author} {\bibinfo {author} {\bibfnamefont {Y.}~\bibnamefont
  {Masubuchi}}\ and\ \bibinfo {author} {\bibfnamefont {T.}~\bibnamefont
  {Uneyama}},\ }\href@noop {} {\bibfield  {journal} {\bibinfo  {journal} {Soft
  Matter}\ }\textbf {\bibinfo {volume} {14}},\ \bibinfo {pages} {5986}
  (\bibinfo {year} {2018})}\BibitemShut {NoStop}%
\bibitem [{\citenamefont {{Espa\~{n}ol}}\ and\ \citenamefont
  {Warren}(1995)}]{Espanol-Warren-1995}%
  \BibitemOpen
  \bibfield  {author} {\bibinfo {author} {\bibfnamefont {P.}~\bibnamefont
  {{Espa\~{n}ol}}}\ and\ \bibinfo {author} {\bibfnamefont {P.}~\bibnamefont
  {Warren}},\ }\href@noop {} {\bibfield  {journal} {\bibinfo  {journal}
  {Europhys. Lett.}\ }\textbf {\bibinfo {volume} {30}},\ \bibinfo {pages} {191}
  (\bibinfo {year} {1995})}\BibitemShut {NoStop}%
\bibitem [{\citenamefont {Kinjo}\ and\ \citenamefont
  {Hyodo}(2007)}]{Kinjo-Hyodo-2007}%
  \BibitemOpen
  \bibfield  {author} {\bibinfo {author} {\bibfnamefont {T.}~\bibnamefont
  {Kinjo}}\ and\ \bibinfo {author} {\bibfnamefont {S.}~\bibnamefont {Hyodo}},\
  }\href@noop {} {\bibfield  {journal} {\bibinfo  {journal} {Phys. Rev. E}\
  }\textbf {\bibinfo {volume} {75}},\ \bibinfo {pages} {0511095} (\bibinfo
  {year} {2007})}\BibitemShut {NoStop}%
\bibitem [{\citenamefont {Langeloth}\ \emph {et~al.}(2013)\citenamefont
  {Langeloth}, \citenamefont {Masubuchi}, \citenamefont {B\"{o}hm},\ and\
  \citenamefont
  {M\"{u}ller-Plathe}}]{Langeloth-Masubuchi-Bohm-MullerPlathe-2013}%
  \BibitemOpen
  \bibfield  {author} {\bibinfo {author} {\bibfnamefont {M.}~\bibnamefont
  {Langeloth}}, \bibinfo {author} {\bibfnamefont {Y.}~\bibnamefont
  {Masubuchi}}, \bibinfo {author} {\bibfnamefont {M.~C.}\ \bibnamefont
  {B\"{o}hm}}, \ and\ \bibinfo {author} {\bibfnamefont {F.}~\bibnamefont
  {M\"{u}ller-Plathe}},\ }\href@noop {} {\bibfield  {journal} {\bibinfo
  {journal} {J. Chem. Phys.}\ }\textbf {\bibinfo {volume} {138}},\ \bibinfo
  {pages} {104907} (\bibinfo {year} {2013})}\BibitemShut {NoStop}%
\bibitem [{\citenamefont {Kawasaki}(1973)}]{Kawasaki-1973}%
  \BibitemOpen
  \bibfield  {author} {\bibinfo {author} {\bibfnamefont {K.}~\bibnamefont
  {Kawasaki}},\ }\href@noop {} {\bibfield  {journal} {\bibinfo  {journal} {J.
  Phys. A: Math. Nucl. Gen.}\ }\textbf {\bibinfo {volume} {6}},\ \bibinfo
  {pages} {1289} (\bibinfo {year} {1973})}\BibitemShut {NoStop}%
\end{thebibliography}%

%------------------------------------------------------------------------------
\clearpage

%------------------------------------------------------------------------------
\end{document}